\documentclass{aa}  

\usepackage{graphicx}
\usepackage{txfonts}
\usepackage{lipsum}
\usepackage{subcaption}
\usepackage{lscape} 
\usepackage{placeins}
\usepackage[colorinlistoftodos,prependcaption,textsize=tiny]{todonotes}

\newcommand{\ra}[4]{#1^{\rm h}#2^{\rm m}#3^{\rm s}.#4}
\newcommand{\dec}[4]{#1^{\rm \circ}#2\arcmin#3\farcs#4}

\begin{document}

   \title{A surprisingly large asymmetric ejection from Mira~A}

\author{T. Khouri\inst{1}\thanks{{\it Send offprint requests to T. Khouri}
   \newline \email{theo.khouri@chalmers.se}}, W. H. T. Vlemmings\inst{1}, D. A. Raudales Oseguera\inst{1,2,3}, D. Tafoya\inst{1}, H. Olofsson\inst{1}, C. Paladini\inst{4}
   M. Maercker\inst{1}, M. Saberi\inst{5,6}, P. Gorai\inst{5,6}, T. Danilovich \inst{7}        }

   \institute{Department of Space, Earth and Environment, Chalmers University of Technology, 412 96 Gothenburg, Sweden 
            \and 
            Escuela de Física, Universidad Nacional Aut\'onoma de Honduras, Ciudad Universitaria, Tegucigalpa, Honduras
            \and
            School of Physics and Center for Relativistic Astrophysics, Georgia Institute of Technology, 837 State Street, Atlanta, GA 30332, USA
            \and
            European Southern Observatory, Alonso de C\'ordova, 3107 Vitacura, Santiago, Chile
            \and
            Rosseland Centre for Solar Physics, University of Oslo, PO Box 1029 Blindern, 0315, Oslo, Norway
            \and
            Institute of Theoretical Astrophysics, University of Oslo, PO Box 1029 Blindern, 0315, Oslo, Norway
            \and
            School of Physics \& Astronomy, Monash University, Wellington Road, Clayton 3800, Victoria, Australia}

   \date{Received September 30, 20XX}
 
  \abstract
   {Stars with masses between roughly 1 and 8~$M_\odot$ end their lives on the asymptotic giant branch (AGB), when intense mass loss takes place,
   with major consequences for the chemical evolution of the universe.
   The mechanism responsible for the outflows is generally accepted to be radiation pressure acting on dust grains that form in the dense extended atmospheres
   of AGB stars. Dust formation is enabled, or at least dramatically enhanced, by the action of convection and stellar pulsations. The complex physics underlying
   convection, stellar pulsations, and dust nucleation precludes predicting AGB mass loss from first principles.}
{Our aim was to characterize the recent mass ejections of the AGB star Mira~A using observations of the inner envelope.
In particular, we studied two lobes observed to be expanding away from Mira~A to 
   obtain empirical insights into the mass-ejection process.} 
   {We investigated the evolution of the lobes using images of polarized light obtained at six epochs using the Spectro-Polarimetric High-contrast Exoplanet REsearch (SPHERE)
   on the Very Large Telescope and of molecular emission at two epochs obtained with the Atacama Large Millimeter/submillimeter Array
   (ALMA).
      Six lines of SO$_2$ were used to investigate the excitation temperature and column density of SO$_2$
   in the lobes. We used the $^{13}$CO~$J=3-2$ line and radiative transfer models to constrain the
   column density of the gas, which allowed us to infer the abundances of SO, SO$_2$, AlO, AlF, and PO in the lobes.}
   {While dust seems confined almost exclusively to the edges of the lobes, gas fills the lobes and displays
   higher densities than expected at the observed radial distances based on the large-scale mass-loss rate of Mira~A, with a total gas mass in the lobes of $\sim 2 \times 10^{-5}~M_\odot$.
   We find the expansion of the lobes to be consistent with both a constant velocity (ejection time in 2010 or 2011) or a decelerating expansion (ejection time in 2012).
  If ejection events with a similar magnitude happen periodically,
  we derive periods between 50 and 200~years to account for the mass-loss rate of Mira~A. This periodicity is very uncertain because of the complexity of the circumstellar environment that
  hampers accurate determinations of the mass-loss rate. We find abundances in the lobes of $\sim 1.5 \times 10^{-6}$ and
  $\sim 2.5 \times 10^{-6}$ for SO and SO$_2$, respectively, when accounting for
  radiative transfer effects and of $2\times10^{-10}$, $6.5\times10^{-10}$, and $4\times10^{-7}$ for AlO, AlF, and PO assuming LTE and optically thin emission.
  The strong variation in brightness of the different features identified in the polarized-light images is puzzling. We suggest that an asymmetric stellar radiation
  field preferentially illuminates specific regions of the circumstellar envelope at a given time, producing a lighthouse-like effect.}
{}
\titlerunning{}
\authorrunning{T. Khouri et al.}


   \maketitle
   \nolinenumbers

\section{Introduction}
The asymptotic giant branch (AGB) represents one of the last phases of the evolution of low- and intermediate-mass stars, with initial masses between $\sim 1$ and $\sim 8~M_\odot$.
During the AGB, stars are cool and have sizes that are more than two orders of magnitude larger than on the main sequence, and a structure that consists of a compact dense core surrounded by
a large convective envelope \citep{Herwig2005}. Stellar pulsations excited in the envelope and convective motions cause the atmosphere to be out of hydrostatic equilibrium, and to have
a larger density scale height. In this dense, cool, extended atmosphere, atoms can form molecules and, if the partial pressures are high enough, dust grains. The newly formed grains significantly affect the opacity in the region
by absorbing and scattering radiation. The radiation pressure felt by the dust grains can affect the dynamics of the gas and dust mixture in the extended atmosphere significantly.
If the radiation pressure force per mass element is stronger than the gravitational pull, material is accelerated outward \citep{Hoefner2018}. The outflows that develop in this way have
mass-loss rates roughly between 10$^{-8}$ and 10$^{-4}~M_\odot~{\rm yr}^{-1}$, and determine the fate of the star.

Predicting the mass-loss rate from first principles is not yet possible because of the complex interplay between convection, stellar pulsations, and dust formation and growth.
The dust-formation process is especially difficult to understand in oxygen-rich AGB stars for which the carbon-to-oxygen number ratio is lower than one, and where dust nucleation is still
a matter of debate \citep{Plane2013,Gail2016,Gobrecht2022,Gobrecht2023}.  In these environments, the lack of dust species that are abundant and can survive at high temperatures \citep{Woitke2006} implies that
relatively large grains ($\gtrsim 0.3~\mu$m) are expected to drive the outflow mainly through the scattering of radiation \citep{Hoefner2008}.
Despite the complex context, wind-driving models have made significant advances in the last decades. For instance, calculations of convection and dust growth in
three-dimensional models now allow for the evolution of ejected filaments to be studied in simulations \citep{Hoefner2019,Freytag2023}.

Our empirical knowledge has also improved significantly, thanks to the advance in instrumentation. Specifically, high angular resolution instruments
now allow for the direct study of the region where dust forms and grows and the wind is accelerated. The detection of grains within the expected size range around AGB stars \citep{Norris2012,Ohnaka2017,Khouri2020} using these
techniques with NaCo and the Spectro-Polarimetric High-contrast Exoplanet REsearch (SPHERE) on the Very Large Telescope (VLT) has provided support for this
mass-loss paradigm. In particular, the emerging picture is of an inner circumstellar structure consisting of a gravitationally bound shell containing aluminum oxide grains surrounded by a region where silicate formation happens and the outflow
is triggered \citep{Karovicova2013,Khouri2015,Hoefner2016}.

In addition, the Atacama Large Millimeter/submillimeter Array (ALMA) and near-IR interferometry
allow the dynamics of the gas in the extended atmospheres to be studied in unprecedented detail \citep{Vlemmings2017,Khouri2018,Ohnaka2019,Khouri2024,Ohnaka2024}.
The very high angular resolution achieved in such observations reveals structures imprinted on the stellar photospheres by convective cells that are spatially resolved \citep{Paladini2018,Vlemmings2024}
and hotspots on the extended atmospheres that remain largely unexplained \citep{Vlemmings2015,Vlemmings2017}.
On larger scales, observations of molecular emission have also revealed new features of
ejection and shaping mechanisms of the outflow.
For instance, the clumpy ejections observed in the carbon-rich AGB star
CW~Leo have been hypothesized to be produced by convective motions \citep{Velilla-Prieto2023}, while the complex morphologies on intermediate scales observed in a sample of oxygen-rich AGB stars
were suggested to originate from interactions with companions \citep{Decin2020}. 

In this context, we report observations with ALMA of molecular emission arising from 
the inner circumstellar environment of the AGB star in the binary system $o$~Ceti, Mira~A. These observations
reveal the expansion of gas and dust in lobes that increase in extent from $\sim 10$~au to $\sim 40$~au from Mira~A
between 2015 and 2023.
In particular, we studied in detail two lobes with enhanced molecular emission first reported by \cite{Khouri2018}. We also analyzed light
scattered by dust observed using SPHERE on the VLT that arises from the edges of such lobes.
In Sects.~\ref{sec:obs} and~\ref{sec:obsRes} we present the observations and the results obtained from the analysis of the data, respectively.
In Sect.~\ref{sec:model} we present radiative transfer
models guided by the results given in Sect.~\ref{sec:obsRes}.
In Sect.~\ref{sec:disc} we discuss our results, while our conclusions are summarized in Sect.~\ref{sec:conc}.

\section{Observations}
\label{sec:obs}

For the molecular line emission, we used data obtained with ALMA on 2017 November 9 and 2023 July 6 and 14 at high angular resolution in
the context of projects 2017.1.00191.S and 2022.1.01071.S. The spatial resolutions achieved in the observations of these projects are $\sim17$ (2017.1.00191.S) and $\sim15$~milliarcseconds (mas; 2022.1.01071.S).
We also employed lower spatial resolution ($0\farcs2$)
data from project 2018.1.00749.S obtained on 2018 November 26 to investigate whether a significant amount of flux is not recovered by the interferometer
in the 2017 observations.
The data obtained in 2017 were reduced and presented by \cite{Khouri2018}, and we refer to that study for details of the data reduction process.
The datasets from 2018 and 2023 were reduced using the standard ALMA pipeline by the observatory.
We retrieved the reduced cubes through the ALMA archive.
The maximum recoverable scales calculated by the ALMA observatory and listed in the archive are $0\farcs329$, $3\farcs033$, and $0\farcs433$ for the data obtained in 2017, 2018, and 2022, respectively.

   \begin{figure*}[ht!]
        \centering
        \includegraphics[width=\textwidth]{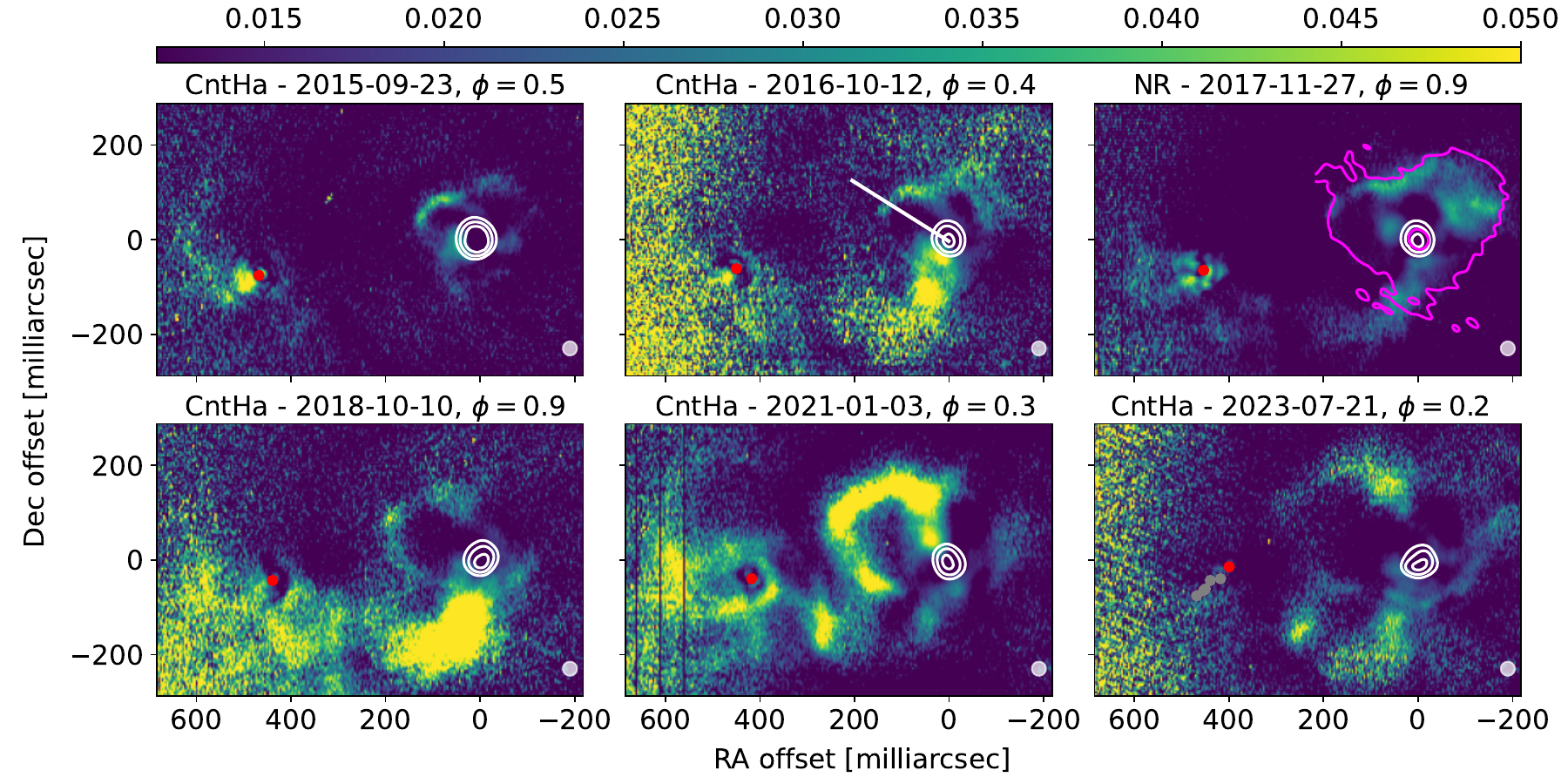}
        \caption{Images of the polarization degree (color map) toward Mira at different epochs in filters CntHa for all epochs except for 2017, which was observed with filter NR.
         The approximate pulsation phases ($\phi$) of Mira~A at the times of the observations are given above each panel.
        The white contours show the total intensity emission from Mira~A at 20\%, 40\%, and 80\% of the peak value.
        The red dots mark the position of the companion in the total intensity images, while the gray dots in the bottom right panel indicate the approximate position
        of the companion in previous epochs. The straight line used to determine the reference point on the NE lobe is shown in the upper middle panel.
        The 3$\sigma$ contour of the SO~$J_K = {8_8-7_7}$ emission in 2017 November (shown in Fig.~\ref{fig:SO})
        is marked by the magenta line in comparison to the polarized-light map at the same epoch. The gray ellipse in the bottom right corner of each panel indicates the
        full width at half maximum of the point spread function in the obtained images.}
        \label{fig:pol_Light}
    \end{figure*}

The observations of polarized light were acquired using the ZIMPOL sub-instrument \citep{Schmid2018} of SPHERE on the VLT
in the context of individual projects and guaranteed-time observations. We searched the archive of the European Southern Observatory and
retrieved data obtained using the CntHa filter on 2015 September 23, 2016 October 12, 2018 October 10, 2021 January 3, and 2023 July 21 and the NR filter on 2017 November 27.
The data were reduced using the pipeline developed at ETH Zurich including bias subtraction, flat-field correction and modulation-demodulation-efficiency correction \citep{Schmid2018}, with
the adoption of values for the modulation-demodulation efficiency of 0.8 when the fast read-out mode was employed (2016 October 12, 2017 November 27, 2021 January 3, and 2023 July 21)
and 0.9 when the slow mode was used (2015 September 9 and 2018 October 10).

The spatial resolution in the ZIMPOL images depends on the performance of the adaptive optics and can be measured using the full width at half maximum (FWHM) of the point spread function (PSF).
The performance is strongly influenced by the atmospheric conditions and the brightness of the source, and can produce FWHM as narrow as $\sim 15$~mas in optimal cases \citep{Schmid2018}.
When a reference point source was observed close in time to the images of the Mira system (two epochs), the reference images were used to determine the FWHM of the PSF.
Alternatively, Mira~B can be used as a PSF reference even if it has a
low brightness compared to Mira~A. In Table~\ref{tab:ZIMPOL}, we give the measurement of the FWHM of Mira~B and the other PSF reference (del Cet), when available. The 
resolution in the SPHERE images presented by us varies between $\sim 19$~mas and $\sim 28$~mas.

\begin{table}[ht!]
\centering 
\caption{\label{tab:ZIMPOL}Stellar pulsation phase at the time of the ZIMPOL observations and spatial resolution in the obtained images.}                 
\begin{tabular}{@{}c | c @{ }c @{\protect\phantom{i} }c @{ }c}
Date & Phase & V mag & PSF Reference & Mira~B \\
 & & & [mas (pixels)] & [mas (pixels)] \\
\hline
2015-09-23 & 0.5 \phantom{(+1)} & 7.9 & -- & 19.8 (5.5) \\
2016-07-21 & 0.4 (+1)& 7.2 & -- & 19.1 (5.3) \\
2017-11-27 & 0.9 (+2) & 8.1 & 22.0 (6.1) & 22.7 (6.3) \\
2018-10-19 & 0.9 (+3) & 8.4 & -- & 22.3 (6.2)\\
2021-01-03 & 0.3 (+6) & 6.4 & 22.7 (6.3) & 24.5 (6.8) \\
2023-07-21 & 0.2 (+9) & 4.5  & -- & 27.7 (7.7) \\
\end{tabular}
\tablefoot{Maximum and minimum stellar light happen at phases 0 and $\sim 0.8$, respectively.}
\end{table}

All images are shown with coordinates relative to the position of Mira~A ($\alpha = \ra{02}{19}{20}{79}$ and $\delta =\dec{-02}{58}{39}{496}$).
The moment-zero maps of the lines shown below are obtained by averaging the channels that show emission. Throughout the paper, we consider a
systemic velocity of Mira~A with respect to the local standard of rest of $\upsilon_{\rm LSR} = 47$~km~s$^{-1}$. The pulsation period of Mira~A
is $\sim 332$~days \citep{Samus2017} and we use the maximum of the light curve around May 5, 2015 as the zero-phase reference.

\section{Observational results}
\label{sec:obsRes}

\begin{figure}
        \centering
        \includegraphics[width=0.49\textwidth]{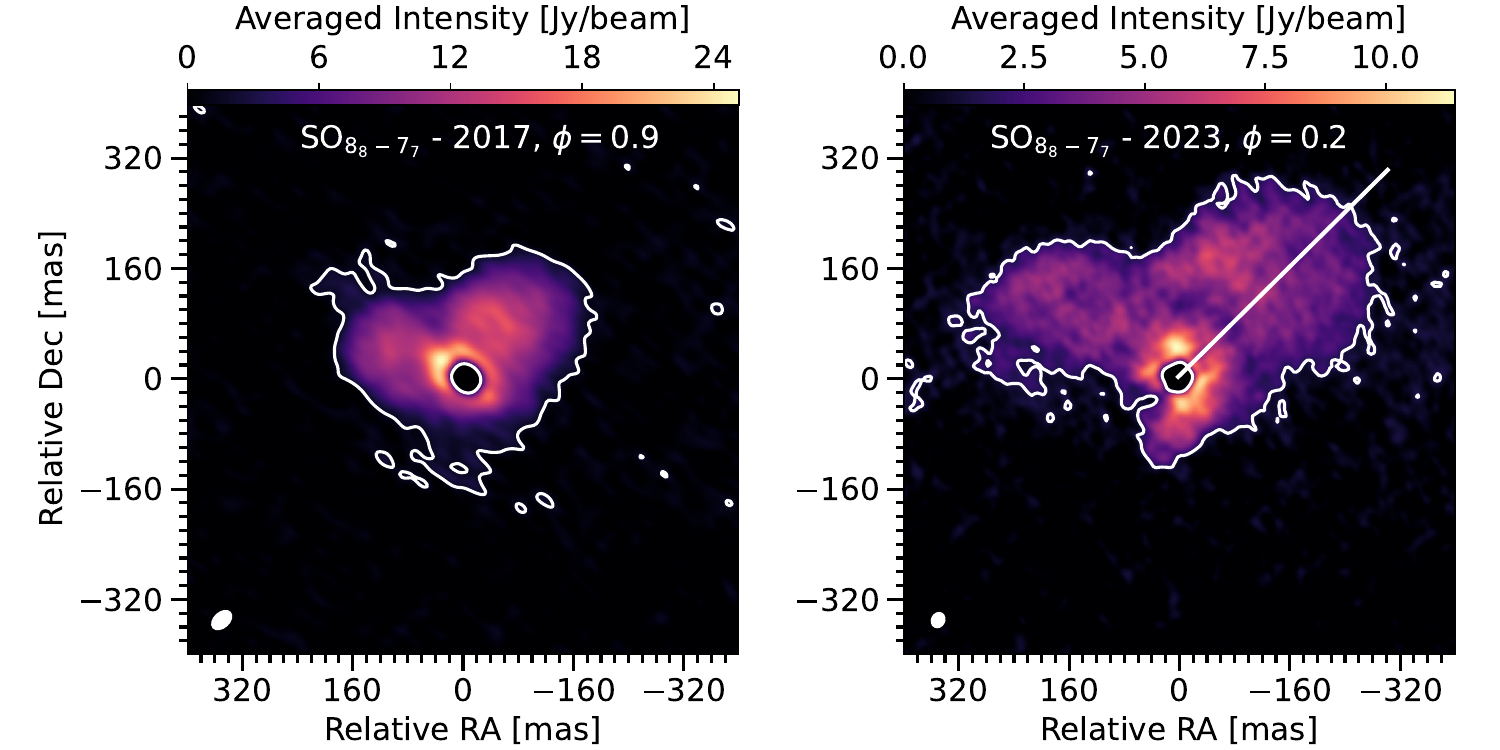}
        \caption{Moment-zero maps of the emission in the SO~$J_K = 8_8-7_7$ line observed in 2017 November 9 and 2023 July 6 and 14.
        The approximate pulsation phases ($\phi$) of Mira~A at the times of the observations are given at the top of each panel.
        The white contours mark where emission reaches three
        times the RMS noise level in each image. The straight white line in the right panel indicates the direction used to measure the size of the NW lobe in the two epochs.}
        \label{fig:SO}
    \end{figure}

A complex structure with two prominent lobes is observed in both the polarized light images (Fig.~\ref{fig:pol_Light}) and the maps of molecular line emission (Figs.~\ref{fig:SO} and additional
maps throughout the paper).
In addition to the lobes, a feature to the south of Mira~A that seems to curve to the east toward Mira~B can be seen in the polarized-light images,
but it has at most a weak counterpart in the molecular line data, as seen in the SO emission to the south of Mira~A (Fig~\ref{fig:SO}).
The lobes are by far the most conspicuous structures in the molecular line maps.
The presence of these lobes were reported by \cite{Khouri2018} based on single-epoch observations (data from 2017 also included in this study). The authors noted that the dust producing the scattered light
seems to be confined mostly to the edge of the molecular emission lobes.
When the observations at other epochs are considered, the evolution of the lobes becomes clear, with expansion visible both in the polarized light images
(Fig.~\ref{fig:pol_Light}) and in the molecular emission maps (Fig.~\ref{fig:SO}).

\subsection{Scattered light}

The polarized light images shown in Fig.~\ref{fig:pol_Light} reveal a complex dust distribution in the system. In particular, the edges of the lobes are traced.
The images acquired in 2015 and 2021 reveal that dust in the lobe to the north-east (NE lobe) of Mira~A is mostly confined to a shell.
The images show the NE lobe to expand significantly over the timespan of the observations we report, $\sim 8$~years.
A lobe to the north-west (NW lobe) is not seen as clearly at all epochs. We speculate that this is caused by the different orientation of the two lobes.
Given that the NW lobe is too faint in most epochs for its morphological evolution to be assessed, we only use the polarized light images to quantify the
expansion of the NE lobe.

The size of the NE lobe at each epoch was measured using
the position of a reference point relative to the center of Mira~A. The reference point was chosen as the intersection of the centre of the filament
with a straight line making a 30$^\circ$ angle with the west-east direction and passing through the center of Mira~A (see Fig.~\ref{fig:pol_Light}). We consider the measurement uncertainty to be half of the width
of the filament at a given epoch, or a larger width from a previous epoch. This was motivated by the facts that the surface brightness distribution of the filament and its width vary significantly as a function of time,
which is probably caused by the varying strength of the signal from the filament between the different epochs, rather than an actual variation in its width. Independently of the cause, this hampers
a more precise identification of a reference point.
The obtained radial positions as a function of time are shown in Fig.~\ref{fig:expansion}.

Using the function \textsc{curve\_fit} of the SciPy Python module, we fitted two functions to the measured points,
a straight line representing constant expansion and a ballistic trajectory assuming a mass of 1.2~$M_\odot$ for Mira~A.
For the constant-expansion model, we obtain the expansion velocity of the NE lobe on the plane of the sky, $\upsilon^{\rm e}_{\rm exp}$,
and its corresponding projected radial extent at the reference epoch (Sep 23, 2015), $r^{\rm e}_\circ$.  
We find $\upsilon^{\rm e}_{\rm exp} =10.5 \pm 1.6$~km~s$^{-1}$ and $r^{\rm e}_\circ = 12.3 \pm 0.8$~au.
Extrapolating backward implies that the NE lobe would have been located at Mira~A's surface $1650\pm200$ days
before the first SPHERE observations (Sep 23, 2015), suggesting an ejection date between late 2010 and early 2011.
For the ballistic model, the two free parameters are the ejection velocity, $\upsilon^{\rm e}_{\rm eje}$, and the ejection time, $t^{\rm e}_{\rm eje}$.
We obtain  $\upsilon^{\rm e}_{\rm eje} = 31.6 \pm 0.6$~km~s$^{-1}$ and $t^{\rm e}_{\rm eje} = 980 \pm 150$~days before September 23, 2015, which corresponds to an ejection time in late 2012 or early 2013.
As shown in Fig.~\ref{fig:expansion}, both models fit the observations well, preventing us from precisely concluding what the initial velocity and time of the ejection of the NE lobe were.

\subsection{Molecular emission in the lobes}

The molecular emission traces the two lobes clearly both in November 2017 and in July 2023. However, the size of the lobes in 2023 is larger than the maximum recoverable scale of the ALMA array used for the observations,
which likely leads to a significant loss of flux. In the observations acquired in 2017, several lines reveal the lobes, as noted by \cite{Khouri2018}.
To characterize the emission in the lobes, we use 9 lines (Table~\ref{tab:lines}) from four different molecules: CO, $^{13}$CO, SO, and SO$_2$. Additionally, we use one line of AlO, AlF, and PO to estimate
the abundances of these molecules in the lobes.
Throughout the text, we do not specify the atomic mass and the vibrational state when referring to the main isotopes of each atom and to transitions within the ground vibrational state, respectively.

A particularly good tracer of the lobe is the
SO~$J_K = {8_8-7_7}$ line (Fig.~\ref{fig:SO}), because this line is optically thick in the lobes, as discussed below. Intrinsically stronger lines, such as $^{29}$SiO~$J=8-7$, show that lower-density gas
surrounds the lobes. As discussed in Sect.~\ref{sec:recentMassLoss}, we estimate the gas surrounding the lobes to have a density about one order of magnitude lower if the excitation temperatures and abundances in the two regions
are not significantly different.
In the second epoch, emission in the SO~$J_K = {8_8-7_7}$ line becomes weaker relative to the
noise level, which remains basically the same at the two epochs, at $\sim 0.5$~mJy/beam. This is caused both by an intrinsic decrease in the surface brightness with the expansion of the lobes and
an observational effect due to loss of flux (see Sect.~\ref{sec:resolved-out}).
Because of the weaker emission and larger emission region size with respect to the beam in 2023, we use the observations from 2017 to derive the physical properties of the gas in the lobes.

\subsubsection{Expansion of the NW lobe}

To estimate the projected expansion velocity of the NW lobe, $\upsilon^{\rm w}_{\rm exp}$,
in a similar way as done for the NE one, we measured the position of its edge using a straight line in the north-west direction passing through
the center of Mira~A (Fig.~\ref{fig:SO}). Our reference point is defined by the interception of this line with the edge of the molecular lobe, which
we define as the contour at three times the root-mean square (RMS) noise level of the molecular emission maps. We use both the
SO~$J_K = {8_8-7_7}$ and $^{13}$CO~$J=3-2$ emission maps averaged over the line profiles. For the images from 2017, the two lines yielded sizes that differed by only 5\%, and we used the average as a representative value.
For the images from 2023, we used only the SO~$J_K = {8_8-7_7}$ maps because the emission of
the $^{13}$CO~$J=3-2$ line was weak compared to the RMS noise level and did not produce continuous contours.
With this approach, we obtain the two magenta points shown in Fig.~\ref{fig:expansion}. A fit of a straight line (constant velocity) implies $\upsilon^{\rm w}_{\rm exp} \sim 12.4$~km~s$^{-1}$ and
$r^{\rm w}_\circ \sim 14.9$~au.
Hence, the edge of the NW lobe would have been at the surface of Mira~A $\sim 1785$ days before the reference epoch
(Sep 23, 2015), which corresponds to an ejection time around late 2010.
Considering a ballistic trajectory, we find ejection velocity, $\upsilon^{\rm w}_{\rm eje}$, and the ejection time, $t^{\rm w}_{\rm eje}$, equal to $\sim 33$~km~s$^{-1}$ and 1090 days before 23 September, 2015, respectively,
implying an ejection time in mid or late 2012.

The values for the NW lobe are more uncertain than those for the NE one, because
of the smaller number of measurements and the less robust method for measuring a reference point at the edge of the lobe.

\begin{table}[ht!]
\caption{Properties of the lines used to study the lobes.}
\label{tab:lines}
\centering 
\begin{tabular}{@{}c @{\protect\phantom{}}l @{\protect\phantom{a}}c @{\protect\phantom{aa}}c @{\protect\phantom{aa}}c @{\protect\phantom{a}}c@{}}
\hline\hline
Mol. & ~Line & $\upsilon$ & g$_{\rm u}$ & A$_{\rm ul}$ & E$_{\rm u}$/k$_{\rm B}$ \\    
& & [GHz] & & [s$^{-1}$] & [K]\\
\hline
\phantom{$^{13}$}CO\phantom{$^\dag$} &~$v=1, J=3-2$ & 342.648 & 7 & $1.45 (-6)$ & 3117\\
$^{13}$CO\phantom{$^\dag$} &~$J=3-2$ & 330.588 & 7 & $1.10 (-6)$ & 31.7\\
\phantom{$^{13}$}AlO\phantom{$^\dag$} &~$N=9-8$ & 344.453& 114 & $2.37 (-3)$ & 82.7 \\
\phantom{$^{13}$}AlF\phantom{$^\dag$} &~$J=10-9$ & 329.642 & 21 & $4.12 (-4)$ & 87.0\\
\phantom{$^{13}$}PO\phantom{$^\dag$} &~$^2\Pi_{3/2}~J, F= 8,8 - 7,7$ & 329.558 & 17 & $6.56 (-4)$ & 386.7\\
\phantom{$^{13}$}PO\phantom{$^\dag$} &~$^2\Pi_{3/2}~J, F= 8,7 - 7,6$ & 329.571 & 15 & $6.56 (-4)$ & 386.7\\
\phantom{$^{13}$}SO\phantom{$^\dag$} &~$J_K = 8_8 - 7_7 $ & 344.311 & 17 & $5.19 (-4)$ & 87.5\\
\phantom{$^{13}$}SO$_2$ &~$J_{K_{\rm a}, K_{\rm c}} = 4_{3, 1}- 3_{2, 2}$ & 332.505 & 9 & $3.29(-4)$ & 31.3 \\
   &~$J_{K_{\rm a}, K_{\rm c}} = 21_{2,20}-21_{1,21}$ & 332.091 & 43 & $1.51(-4)$ & 219.5 \\
&~$J_{K_{\rm a}, K_{\rm c}} = 34_{3,31}-34_{2,32}$ & 342.762 & 69 & $3.45(-4)$ & 581.9 \\
&~$J_{K_{\rm a}, K_{\rm c}} = 36_{5,31}-36_{4,32}$ & 341.674 & 73 & $4.34 (-4)$ & 678.5 \\
&~$J_{K_{\rm a}, K_{\rm c}} = 40_{4,36}-40_{3,37}$ & 341.403 & 81 & $4.12 (-4)$ & 808.4 \\
&~$J_{K_{\rm a}, K_{\rm c}} = 44_{5,39}-44_{4,40}$ & 329.689 & 89 & $4.33 (-4)$ & 984.9 \\
\hline
\end{tabular}
\end{table}

\subsubsection{Resolved-out emission}
\label{sec:resolved-out}

To estimate whether a significant amount of flux was resolved out in the observations of the low-excitation lines acquired in 2017, we used the products available in the ALMA archive
from project 2018.1.00749.S. These data were acquired on 2018 November 26, which is close to one year
after the observations we analyze in this study and cover the lines $^{13}$CO~$J= 3-2$, SO $J_K = 8_8-7_7$, AlO~$N= 9-8$ and SO$_2 ~J_{K_{\rm a}, K_{\rm c}} = 4_{3, 1}- 3_{2, 2}$, 
$21_{2,20}-21_{1,21}$, and $34_{3,31}-34_{2,32}$. The achieved spatial resolution is significantly lower, with a beam size of $0\farcs25 \times 0\farcs19$ and a position angle
varying between -71$^\circ$ and -77$^\circ$.

To compare both datasets, we use the \textsc{imsmooth} function in CASA to smooth
the higher-resolution image cubes from 2017 to the same resolution as the 2018 ones, using the appropriate beam size and orientation at the frequency of each line.
Then, we extract spectra from both datasets using a circular region with diameter $0\farcs4$ centered on the continuum peak.
As shown in Fig.~\ref{fig:resOutFlux}, the lines have very similar strengths in both epochs. The only line displaying strong variation
is AlO~$N=9-8$, which is almost a factor of two weaker in 2018. This could be caused by dissociation or a change in the excitation of the molecule.
Based on this comparison, we conclude that there is no significant loss of flux in the observations from 2017 despite the relatively small beam.

\subsubsection{SO$_2$ population diagrams}

We identify six relatively strong lines of SO$_2$ (Table~\ref{tab:lines}) in the spectra of the data obtained in November 2017. The energies of
the upper level of these lines range between $\sim 31$ and $\sim 985$~K, covering the expected gas temperatures
in the region probed. Hence, these are excellent for tracing the excitation temperature, $T_{\rm SO_2}^{\rm exc}$, and column density, $N_{\rm SO_2}$, of SO$_2$ molecules
in the lobes.

To retrieve the excitation temperature and column density of SO$_2$ per pixel, we averaged the emission in these six SO$_2$ lines over frequency to obtain the so-called moment-zero maps of the lines. These maps were
then used to obtain a population diagram \citep{Goldsmith1999} for each pixel. To calculate the number of molecules in the upper level of each transition at each given pixel ($N_{\rm u}$), we used the expression
\begin{equation}
\label{eq:Nu}
N_{\rm u} = 4~\pi~d^2~F~l^2_{\rm pxl}~h~\nu~A_{\rm ul},
\end{equation}
where $d$ is the distance to the Mira system \citep[$\sim 100$~pc,][]{vanLeeuwen2007,Haniff1995}, $F$ is the flux of the given line in the given pixel, $l_{\rm pxl}$ is the size at the distance of Mira corresponding to the angular size of the pixels,
$h$ is the Planck constant, and $\nu$ and $A_{\rm ul}$ are the frequency and the Einstein-A coefficient
of the relevant transition, respectively. The flux per pixel was obtained by dividing the flux per beam by the ratio between beam and pixel areas
at the frequency of each line. The pixel size in the images produced by us corresponds to 3.6~mas, or 0.36~au at 100~pc, at all frequencies.

   \begin{figure}
        \centering
        \includegraphics[width=0.45\textwidth]{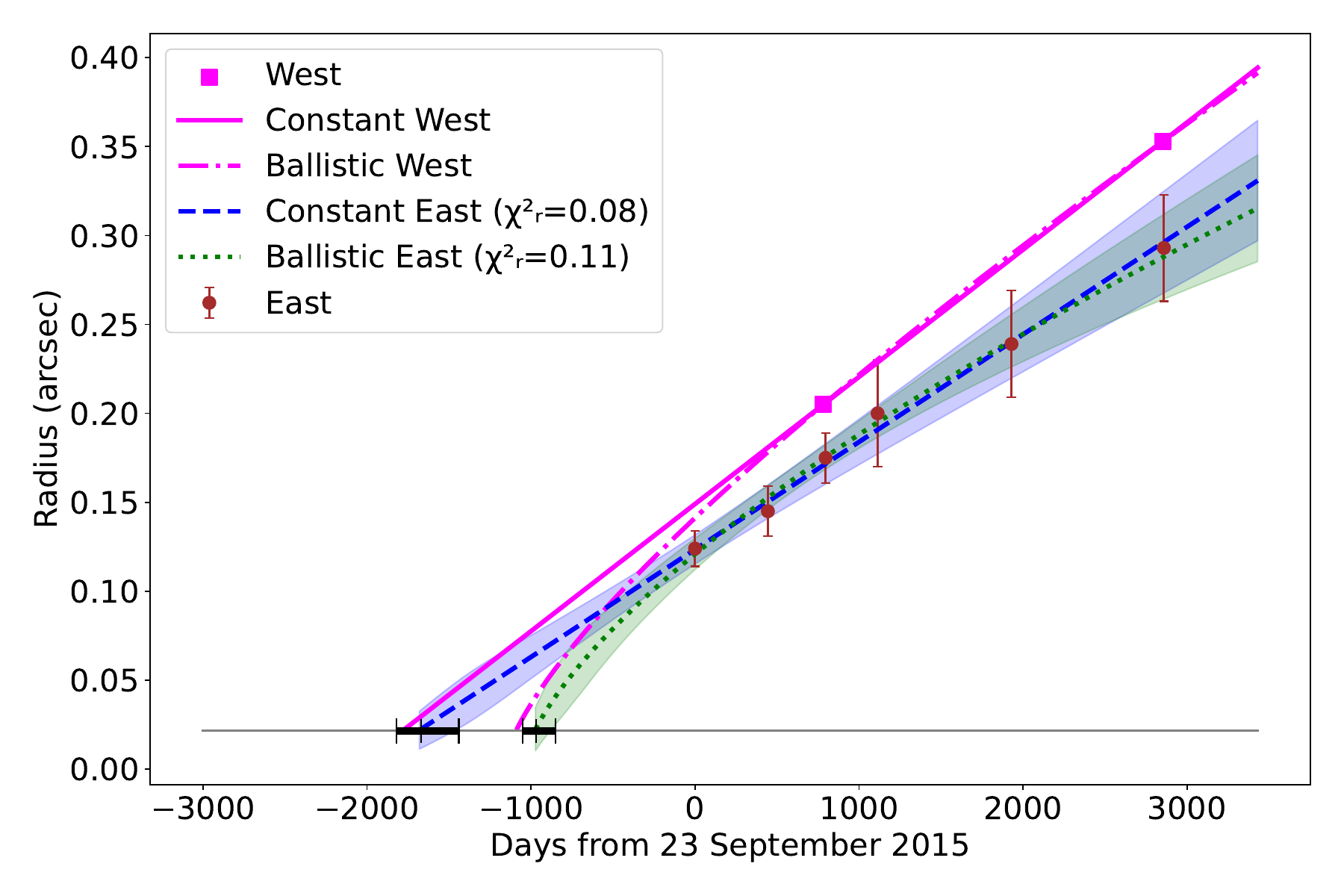}
        \caption{Observed distance on the plane of the sky of the reference point of the NE (maroon circles) and NW (magenta squares) lobes.
        The magenta solid line and blue dashed line show the fits to the observed extents of the NW and NE lobes, respectively, assuming constant expansion. The magenta dot-dashed and green dotted
        line shows a fit to the extent of the NW and NE lobes, respectively, assuming a ballistic trajectory for a star with 1.2~$M_\odot$. The black intervals mark the uncertainty interval on the ejection times of the
        lobe to the NE based on the two models.}
        \label{fig:expansion}
    \end{figure}

   \begin{figure*}
        \centering
        \includegraphics[width=\textwidth,trim={0 0 3cm 0},clip]{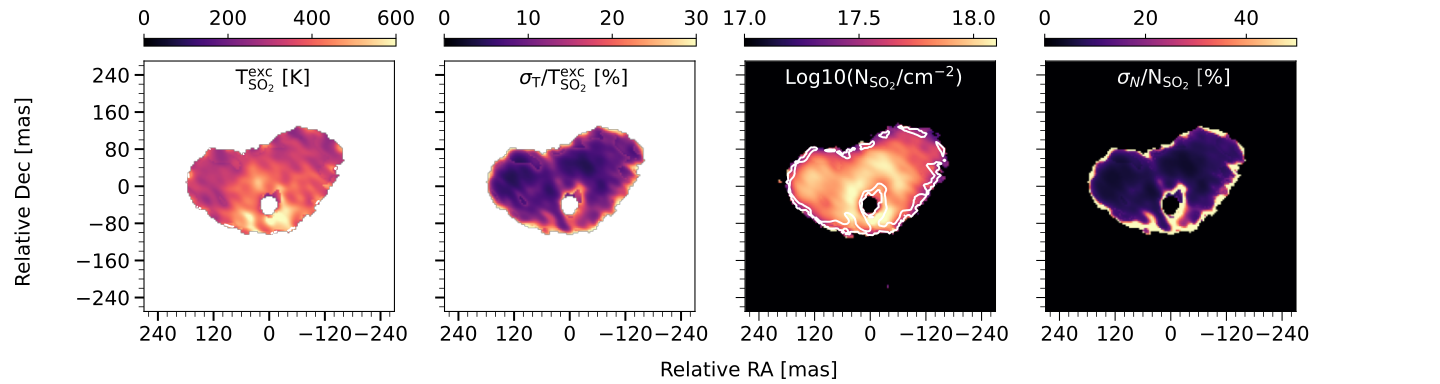}
        \caption{Results from the SO$_2$ populations diagram. From left to right, the panels show the obtained maps of $T_{\rm SO_2}^{\rm exc}$ in Kelvin, the uncertainty on $T_{\rm SO_2}^{\rm exc}$ in percentage,
        log$_{10}(N_{\rm SO_2}~{\rm cm^{-2}}$), and the uncertainty on $N_{\rm SO_2}$ in percentage. The white contours indicate the 30\% uncertainty level on $N_{\rm SO_2}$.}
        \label{fig:SO2_pop}
    \end{figure*}

After obtaining values for the populations of the six levels for each pixel,
we derived the excitation temperature and column density, $N$, per pixel by fitting straight lines to the data in the $log\left ( \frac{N_{\rm u}}{g_{\rm u}} \right ) \times E_{\rm u}$ plane,
where $g_{\rm u}$ and $E_{\rm u}$ are the degeneracy
and excitation energies of the upper level, respectively. An example of such a fit is shown in Fig.~\ref{fig:popDiamExam}.
The excitation temperature is equal to minus the inverse of the slope of the fitted line, while the column density
is equal to the partition function at the relevant excitation temperature, $Z(T_{\rm SO_2}^{\rm exc})$, times the exponential of the intercept of the fitted line. We used the partition function as a function of temperature
provided by \cite{Underwood2016} and available through the ExoMol database.

To calculate the uncertainties in the derived quantities, we used a Monte Carlo method
and simulated $3\times10^4$ observations. In each simulation, the line fluxes per pixel were randomly chosen from Gaussian distributions
centered on the values of the given pixel of the moment zero maps and with standard deviations equal to the observed root-mean square noise of the given map.
We do not include the absolute flux calibration uncertainty in the estimate of the excitation temperature because that does not affect the relative strengths of the lines,
which is used to determine the excitation temperature.
We considered in our calculations all pixels for which the fluxes of at least four SO$_2$ lines were higher than three times the root-mean square noise.
The standard deviations on the determined values of $T_{\rm SO_2}^{\rm exc}$ and $N_{\rm SO_2}$ for each pixel were calculated based on the distributions of the $3\times10^4$
values of $T_{\rm SO_2}^{\rm exc}$ and $N_{\rm SO_2}$. The maps of the excitation temperature,
column density, and the corresponding standard deviation are shown in Fig.~\ref{fig:SO2_pop}.
Integrating over the SO$_2$ emission region, we obtain a total number of SO$_2$ molecules $\sim 6.6\times10^{45}$.

Based on the obtained temperature maps and channel-averaged emission of the SO$_2$ lines, we calculated maps of the averaged optical depth for each line (Fig.~\ref{fig:SO2_tau}).
Given that the averaged optical depth of the SO$_2$ lines peaks at $\sim 0.5$, optical depth effects are expected to affect the lines to some extent. Nonetheless, the derived excitation temperature distribution
are also supported by our radiative transfer models discussed below, implying that these transitions provide a good probe of the SO$_2$ excitation in this region.

\subsubsection{Gas column densities}

To translate the observed $N_{\rm SO_2}$ values to gas column densities, the abundance of SO$_2$ needs to be determined. This is not straightforward, however, because sulfur can be atomic, in molecules
(mainly SO or SO$_2$), and even in dust grains. Therefore, there is a large uncertainty on the SO$_2$ abundance. 

   \begin{figure*}
        \centering
        \includegraphics[width=0.8\textwidth]{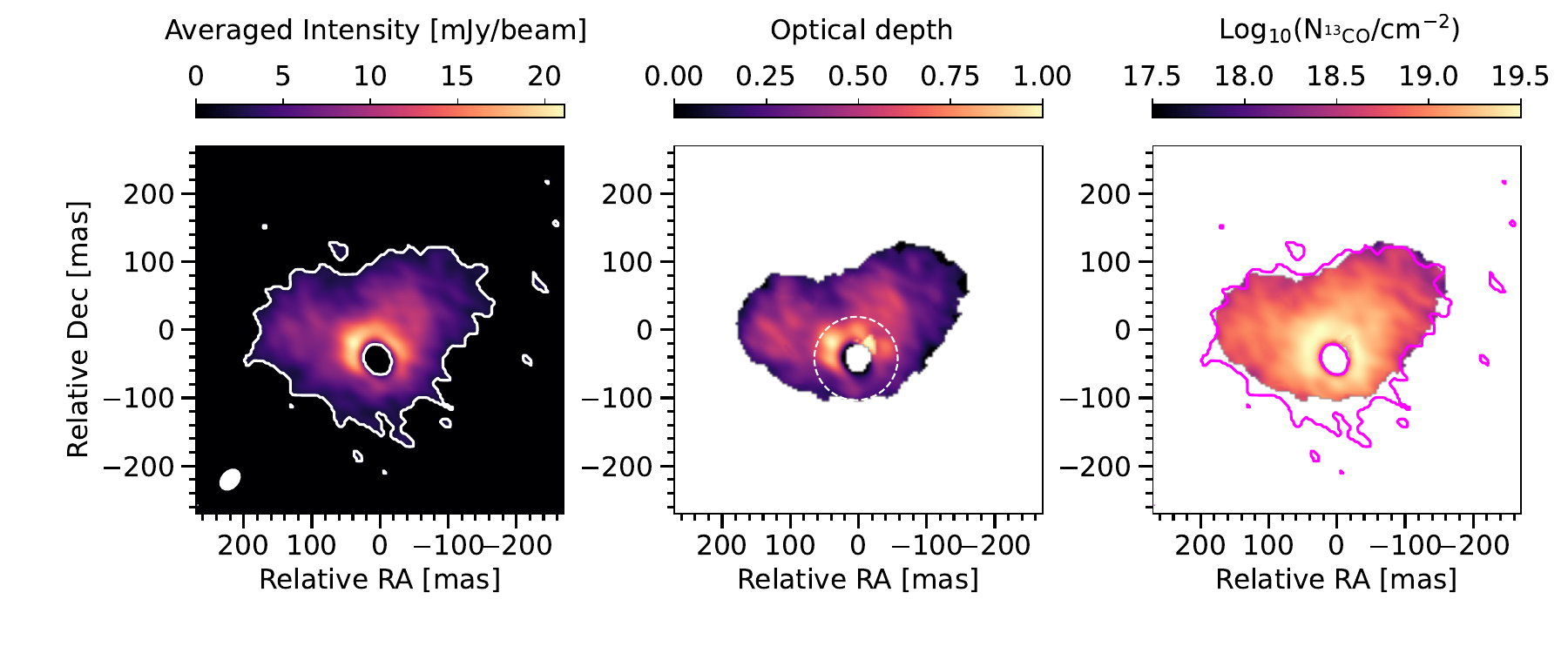}
        \caption{Observed emission (left) and inferred optical depth (middle) of the $^{13}$CO~$J=3-2$ line and corresponding calculated column density of $^{13}$CO (right)
        assuming optically thin emission and the same excitation temperature distribution derived for SO$_2$. The dashed white circle in the middle panel marks the 60~mas radius inside of
        which the derived column densities are not reliable because of high optical depths. The white and magenta contours in the left and right panels, respectively, mark emission equal to three times the RMS noise level.}
        \label{fig:13CO}
    \end{figure*}

In light of this, we use the $^{13}$CO~$J=3-2$
line to estimate the gas column density and derive the SO$_2$ abundance.
The abundance of $^{13}$CO should be much better constrained, because virtually all $^{13}$C is expected to be locked in $^{13}$CO.
To obtain the column density of $^{13}$CO molecules, we assume that the excitation temperature
of the $^{13}$CO~$J=3-2$ line is the same as the one derived from the SO$_2$ line. The simplest scenario in which this would be true is if all the lines in question are excited under LTE conditions.
Our population diagram calculations for SO$_2$ suggest that this is a likely possibility, given that a single excitation temperature produced very good fits to each pixel in both lobes. Moreover, the derived temperatures vary smoothly over the
lobes. The relatively low value
of the Einstein-A coefficient of the $^{13}$CO~$J=3-2$ line implies that collisions dominate the excitation for the gas densities we infer.
The average optical depth of the $^{13}$CO~$J=3-2$ line in the lobes (Fig.~\ref{fig:13CO}) implies that the estimated column densities are a lower limit because of
opacity effects. Nonetheless, we do not expect a strong effect in the lobes. We address the problem of the optical depths by calculating radiative transfer models below.

Under the assumption of optically thin emission and excitation temperature equal to that of SO$_2$, we obtain the column density map shown in Fig.~\ref{fig:13CO}.
As can be seen, the highest optical depths in the $^{13}$CO~$J=3-2$ line are reached in the innermost regions of the envelope, where densities are high.
Hence, we expect strong optical depth effects to affect the measured column density of $^{13}$CO for regions at radii $r < 60$~mas from the center of Mira~A.

\subsubsection{Estimated molecular abundances}

Assuming a $^{13}$CO abundance relative to H$_2$ of $3 \times 10^{-5}$ \citep{Hinkle2016,Khouri2018}, our results imply an average abundance of SO$_2$ with respect to H$_2$ of  $\sim 2.5 \times 10^{-6}$ for $r > 60$~mas from the center of Mira~A.
As a reference, we consider the solar sulfur abundance with respect to H$_2$ \citep[$\sim 2.6 \times 10^{-5}$,][]{Asplund2021}
to be an upper limit to possible values of sulfur-bearing molecules.
Similarly, we estimated the abundances of other molecules with strong line emission in the lobes using the lines
SO~$8_8-7_7$ (Fig.~\ref{fig:SO_tau}), AlF~$J=10-9$ (Fig.~\ref{fig:AlF}), AlO~$J=9-8$ (Fig.~\ref{fig:AlO})  and PO~$^2\Pi_{3/2}~J= 8 - 7$ (Fig.~\ref{fig:PO}) and
the excitation temperatures and gas column densities derived above using the SO$_2$ and $^{13}$CO lines.
As before, we consider only the emission arising from $r > 60$~mas. The partition function and transition parameters adopted from the literature for each molecule are specified in Appendix~\ref{sec:transParam}.
We find average abundances of $\sim 4.5\times10^{-7}$ for SO, of $6.5\times10^{-10}$ for AlF, of $2\times10^{-10}$ for AlO, and $4\times10^{-7}$ for PO.
While emission in most lines is overall optically thin, the SO~$J_K = 8_8 - 7_7$ line reaches high optical depths in the lobes, implying that the abundance derived in this way represents only a lower limit.
The radiative transfer calculations discussed below provide a better estimate of the abundance.
The $J=8-7$ line of $^{29}$SiO was also inspected by us, but the optical depths in this line are so high that estimating an abundance associated with the emission has little meaning.

\section{Radiative transfer models for the molecular lines}
\label{sec:model}

To better constrain the morphology, abundances, and density and velocity distributions of the gas in the lobes, we calculated radiative transfer models using the code LIME \citep{Brinch2010}.
 The model output consists of a line emission cube for each line. These cubes were convolved to the resolution of the observations at the central frequency of each transition.
 The stellar continuum was subtracted to produce continuum-free model line cubes. The modeled spectra were extracted using circular apertures centered on the model star and
 with radii equal to 40~mas, 100~mas, and 200~mas. These were compared to spectra extracted from the same regions from the observed data (Fig.~\ref{fig:lineFits}). 

\subsection{Model setup}

We base the structure of the model on the results presented in Sect.~\ref{sec:obsRes} and on the model for $r < 84$~mas around Mira~A presented by \cite{Khouri2018}.
The stellar temperature ($T_\star = 2100$~K) and radius of the central star at 338~GHz, $R_\star^{\rm 338~GHz} = 21.25$~mas (or 2.125~au at 100~pc), were adopted from \cite{Vlemmings2019}.
Since LIME does not include a central star explicitly, we include a central blackbody consisting of a dust-rich sphere that is optically thick at all wavelengths considered in the calculation and has a
temperature equal to the brightness temperature of the central star at 338~GHz (2100~K).

For CO, rotational levels up to J= 40 in the
ground, first, and second vibrational states are included. The collisional rates
we use were obtained by \cite{Castro2017}.
The CO and $^{13}$CO abundances relative to H$_2$ were kept fixed in the models at $4 \times 10^{-4}$ and $3 \times 10^{-5}$, respectively, following \citep{Khouri2018}, while the abundances of SO and SO$_2$
were allowed to vary to fit the observed line strengths. In this way, the gas density in the lobes is constrained mostly by fitting the strength of the $^{13}$CO line extracted using different apertures, while the abundances of SO and SO$_2$
are the most relevant parameters to match the strength of the lines of each species. The relative strengths of the SO$_2$ lines together with the radial brightness distribution of each line
are the main constraints for the gas temperature distribution.

\subsubsection{Envelope physical structure}

\begin{figure}
        \centering
        \includegraphics[width=0.45\textwidth]{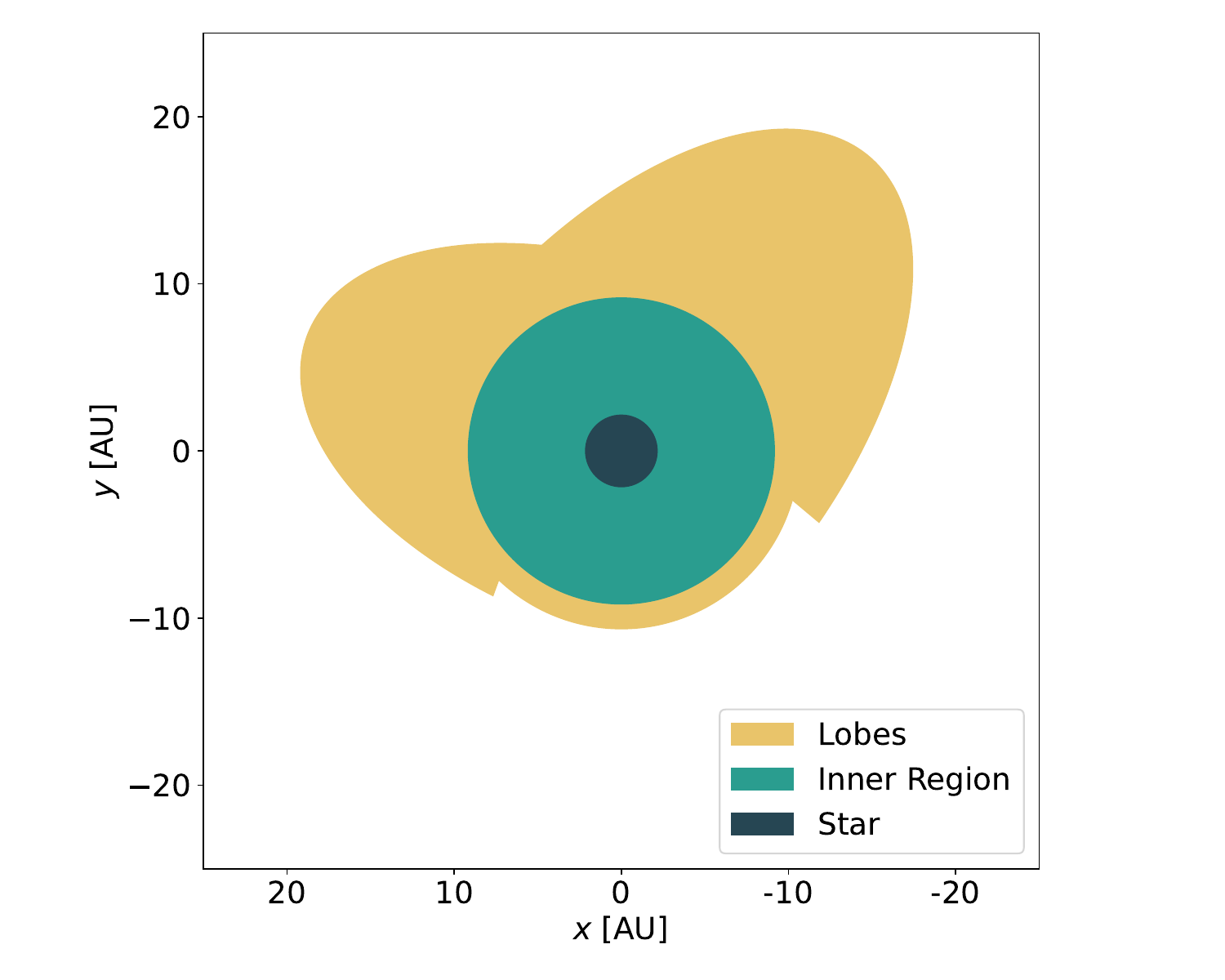}
        \caption{Schematic view of the model with the different regions indicated. This view is a cut through the plane defined by the two major axes of the ellipses (main plane of the model).
        For the ray-tracing, the model is rotated about the major axis of
        the smaller lobe by an angle of 20$^\circ$. In this way, the larger lobe is inclined toward the observer to produce the observed blueshifted emission.}
        \label{fig:modelSchematic}
    \end{figure}

To approximate the structure of the inner circumstellar environment of Mira~A, we consider a sphere (Inner Region) and two lobes described by truncated ellipsoids (Fig.~\ref{fig:modelSchematic}).
The radius of transition between the Inner
Region and the lobes, $R_{\rm trans}$, was set to 91.375~mas (4.3~$R_\star$)
based on the size of the bright ring seen in the $^{13}$CO~$J=3-2$ channel maps (Fig.~\ref{fig:13CO-channels}). The lobes are described by ellipsoids centered on
Mira~A and truncated at the intersection with a plane perpendicular to the major axis of the given lobe and located $2~R_\star$ away from the center of Mira~A. This approach was chosen
to avoid the lobes extending in front of the star because that would produce blueshifted absorption against the stellar continuum which is not observed in the spectral lines. The lobes envelop
the Inner Region for radii smaller than $4.3~R_\star$.

   \begin{figure*}[th!]
        \centering
        \includegraphics[width=0.8\textwidth]{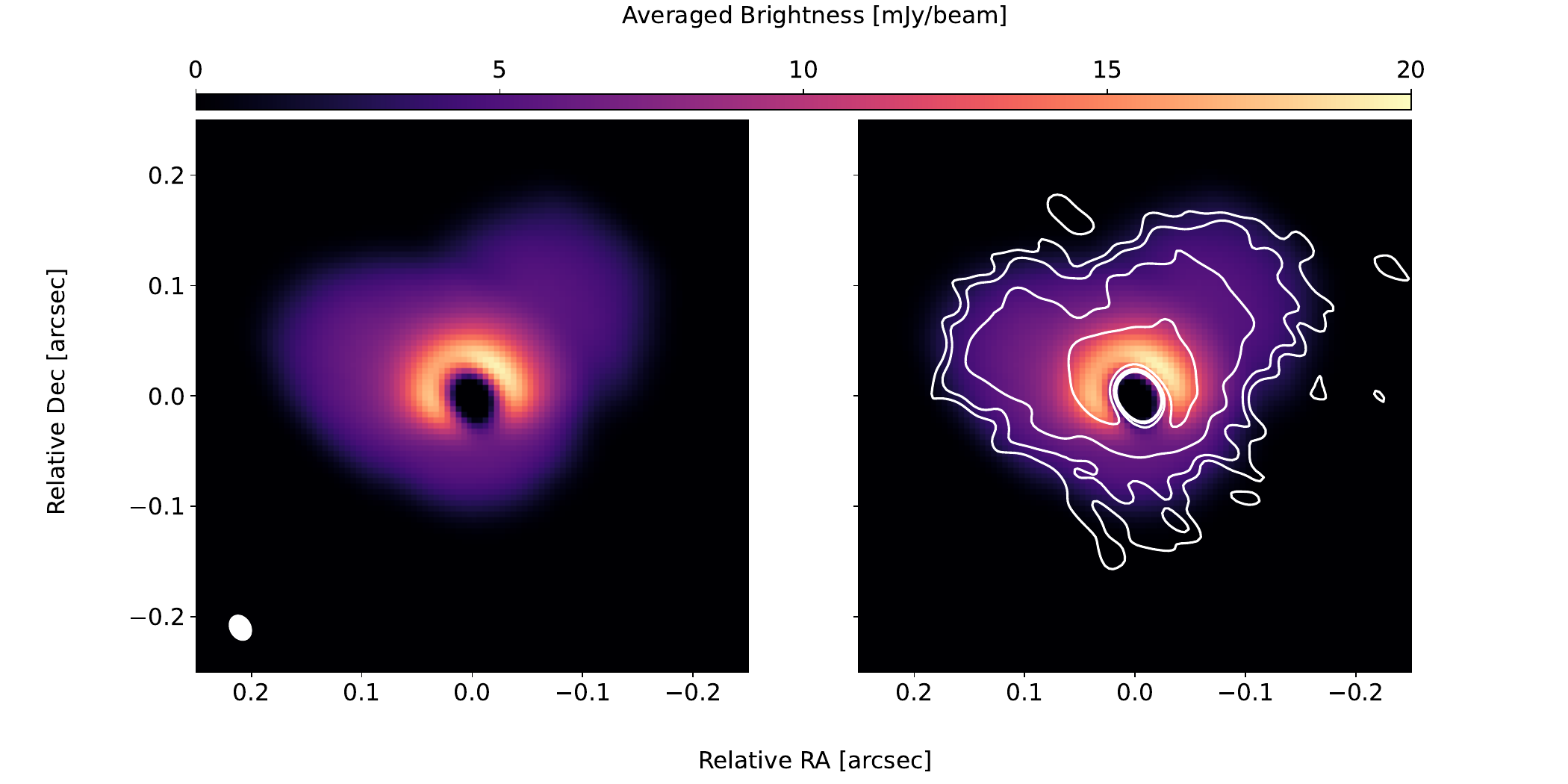}
        \caption{Moment-zero map of the $^{13}$CO~$J=3-2$ line of the best-fitting model (color map) compared to the moment zero of the observed line (white contours at levels of 3, 5, 10, and 20 times the RMS noise).
        The line emission and physical structure of the same model are shown in Fig.~\ref{fig:lineFits} and Fig.~\ref{fig:model}, respectively.}
        \label{fig:modelMoment0}
    \end{figure*}

As a reference, we refer to the plane defined by the two major axes of the ellipsoids as the main plane of the model in the following description.
As shown in Fig.~\ref{fig:modelSchematic}, we oriented the major axis of the NE lobe 20$^\circ$ away from the $x$ direction and toward the $y$ direction on the main plane. For the NW lobe,
the major axis was oriented 5$^\circ$ off the $x=y$ direction toward the $y$ axis on the main plane. This implies an angle of 110$^\circ$ between the two major axes of the lobes.
These parameters were chosen to reproduce the morphology traced by the molecular lines (e.g., Fig.~\ref{fig:modelMoment0}). The lobe to the NW
was inclined toward the observer by an angle of $20^\circ$ by rotating the main plane about the major axis of the NE lobe.
This is necessary to reproduce the observed blueshifted emission in the channel maps (Fig.~\ref{fig:13CO-channels}).

The kinetic temperature, $T$, density, $\rho$, and velocity distributions of the gas were assumed to be power-law functions of the radial distance to the star.
The gas density and temperature
decrease radially and are described by one power-law function within the Inner Region and different ones in the lobes. The steepness and reference values of the
functions were determined from the observations as detailed below.
The gas velocity field has only a radial component, to which a turbulence velocity of 3~km s$^{-1}$ is added.
The lines are also affected by thermal broadening based on the local kinetic temperature.

\subsubsection{Model parameters and modeling approach}
\label{sec:modelParam}

The good description of the level populations of SO$_2$ using a single temperature for each pixel suggests that the obtained
temperature values are good approximations to the average gas kinetic temperature along the line of sight, and that the populations of the corresponding
SO$_2$ levels are excited under LTE.
Based on the obtained map, we
adopt a continuous temperature profile that is equal to the observed continuum temperature at $R^{\rm 338~GHz}_\star$ (2100~K) and
drops to 400~K at $R_{\rm trans} = 4.3~R_\star$ (91.375~mas) and to 300~K at 10.4~$R_\star$ (220~mas). This corresponds to a broken power law with exponents $\sim -1.14$ for $r < R_{\rm trans}$ and $\sim -0.33$ for $r > R_{\rm trans}$.

\begin{figure*}[th!]
        \centering
        \includegraphics[width=0.9\textwidth]{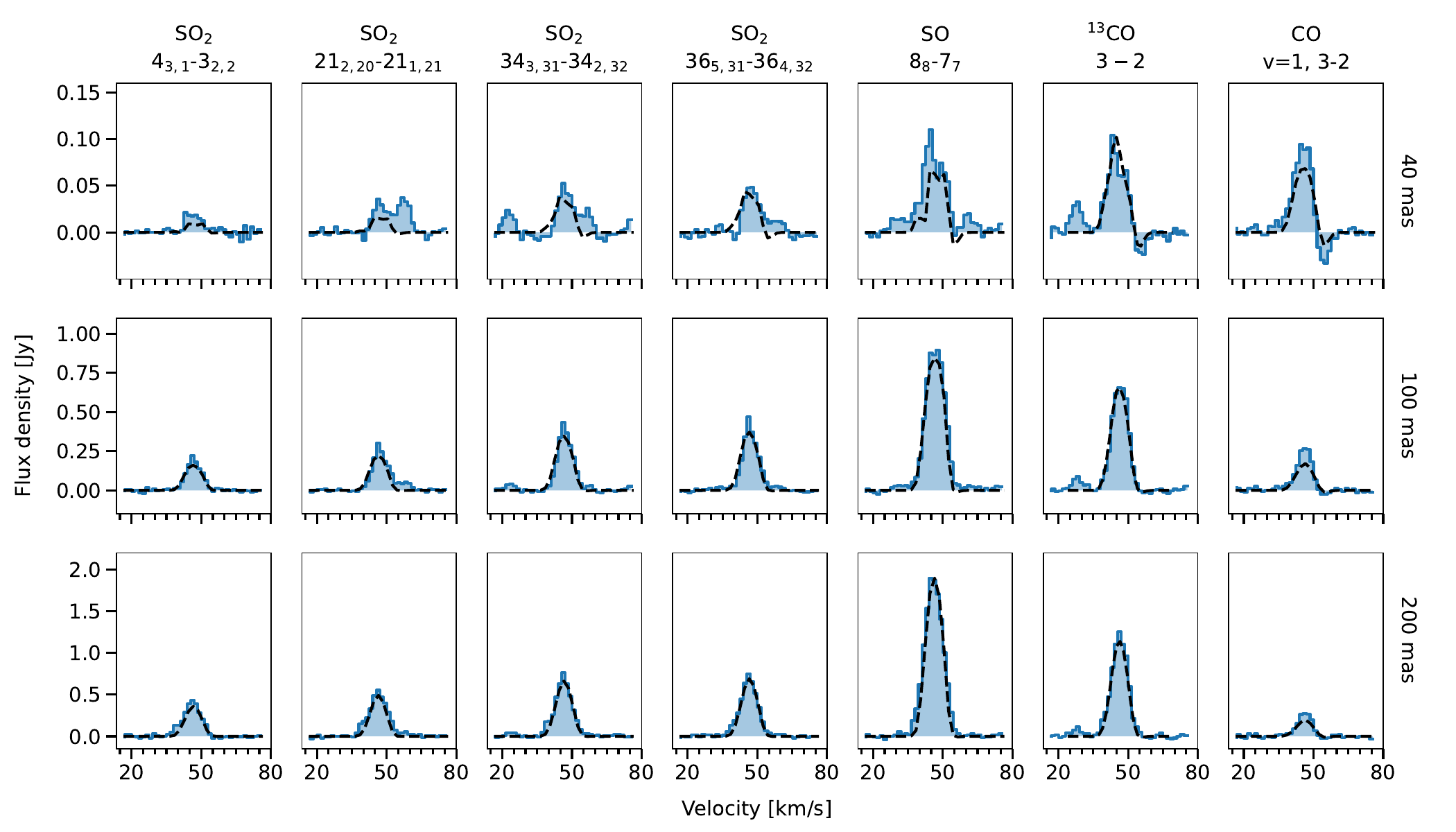}
        \caption{Observed lines extracted from three different apertures (blue histograms)
        compared to the results from the radiative transfer model (dashed black lines) summarized in Table~\ref{tab:model}. A systemic velocity of 47~km~s$^{-1}$ was added to the model.}
        \label{fig:lineFits}
    \end{figure*}

We use the estimated radial decrease of the $^{13}$CO column density to assess the steepness of the gas density profile.
For this, we consider only the regions over which the lobes have roughly constant width (up to $r \sim 180$~mas) to prevent effects from the narrowing of the lobes at the edges.
We find the $^{13}$CO column density to decrease by less than a factor of two from the average value at $r=90$~mas.
This factor should not be strongly affected by the effect of the morphology of the lobes on the observed column densities and, hence, should be
indicative of the steepness of the gas density profile, implying a weak dependence of the gas density with radius in the lobes.
For a power-law distribution $\rho_{\rm gas} = \rho_\circ (r/R_\star^{\rm 338~GHz})^{\eta}$, this implies $\eta \sim -0.75$.
The velocity of the gas in the models is also described by a linear
function of $r$, $\upsilon(r) = \upsilon_{\rm in} + (\upsilon_{\rm out}-\upsilon_{\rm in}) \times (r-R_{\rm in})/(R_{\rm out}-R_{\rm in})$,
where $\upsilon_{\rm in}$ and $\upsilon_{\rm out}$ are the velocities at the inner ($R_{\rm in}$) and outer ($R_{\rm out}$) radii of the given region. Negative and positive velocities
correspond to infall and outflow, respectively.
The velocity distribution was assumed to be on average negative in the inner spherical region and positive in the lobes,
because of the redshifted absorption observed in the line profiles (Fig.~\ref{fig:lineFits}) and of the measured outward expansion of the lobes, respectively.

The density and velocity distributions were varied from our initial model to reproduce the observed line emission.
We compared the model predictions to the observed lines $J_{K_{\rm a},K_{\rm c}} = 4_{( 3, 1)}- 3_{( 2, 2)}$, $21_{( 2,20)}-21_{( 1,21)}$, $34_{( 3,31)}-34_{( 2,32)}$, and $36_{( 5,31)}-36_{( 4,32)}$ of SO$_2$,
$J_K = 8_8-7_7$ of SO, and $J=3-2$ of $^{13}$CO
 considering local thermodynamical equilibrium and from line ${\rm v}=1, J=3-2$ of CO considering non-LTE effects. Although non-LTE effects can be important for
 the CO~${\rm v}=1, J=3-2$ line, it is difficult to assess whether our calculations
 provide a good approximation for the radiative excitation, given the uncertain direct stellar radiation field as discussed below.  
 
\subsection{Model results}
\label{sec:modelRes}

As shown in Figs. \ref{fig:modelMoment0} and ~\ref{fig:lineFits}, the observed lines are fairly well reproduced by a radiative transfer model with the parameters derived in Sect.~\ref{sec:modelParam} based on the population diagram analysis.
The temperature, density and velocity profiles used in the model are summarized
in Table~\ref{tab:model} and shown in Fig.~\ref{fig:model}.
To reproduce the weaker emission observed south of the star, we reduce the gas density in the Inner Region for $y < R_\star$, where $y$ is the coordinate
on the main plane of the model (Fig.~\ref{fig:modelSchematic}), to the value at the base of the lobes ($ 8 \times 10^{14}$ H$_2$~molecules~m$^{-3}$, Table~\ref{tab:model}).
Additionally, the gas densities in the NE lobe are 30\% lower than in the NW lobe at the same radial distance. This difference was introduced because
otherwise the relative brightnesses of the lobe does not fit the observed emission maps.

The density needs to be relatively high ($\sim 10^{17}$~molecules/m$^{3}$) close to the star for the CO~$v=1, J=3-2$ line to be as strong as observed.
A steep density distribution is required to reproduce the lower-excitation $^{13}$CO~$J=3-2$ emission arising from a larger region.
The model contains gas masses of $\sim 8 \times 10^{-5}$~M$_\odot$ and $\sim 2.1\times10^{-5}$~M$_\odot$ in the Inner Region and the lobes, respectively.

We find abundances of SO and SO$_2$ in the lobes of $\sim 1.5 \times 10^{-6}$ and $\sim 2.5 \times 10^{-6}$, respectively. Our calculations also show that the abundance of SO$_2$
in the high gas density region close to the star ($r \lesssim R_{\rm trans}$) is likely significantly lower than in the lobes, with values $\sim 2 \times 10^{-7}$,
because otherwise the lines from
the smallest aperture considered by us are predicted to be stronger than observed as shown in Fig.~\ref{fig:SO+SO2_abundances}. Our models also suggest a similar behavior for the SO abundances, but 
the high optical depths of the SO line in the Inner Region make constraining the SO abundance more difficult.

\begin{table}[ht!]
\caption{Results from the radiative transfer model.}
\label{tab:model}
\centering
\begin{tabular}{l@{}l}
Inner Region ($r < R_{\rm trans}$) \\
Temperature law & $2100~(r/R_\star)^{-1.15}$~K\\
Density law & $ 10^{17}~(r/R_\star)^{-3.2}$~H$_2$~mol.~m$^{-3}$ \\
Total gas mass & $\sim 8 \times 10^{-5}$~M$_\odot$ \\
Velocity law & $-9 + 9\times \left(\frac{r-R\star}{R_{\rm trans} - R_\star}\right)$~km~s$^{-1}$\\
Abundances SO, SO$_2$ & $\sim 2 \times 10^{-7}$, $\sim 2 \times 10^{-7}$\\
\hline
Lobes ($r \geq R_{\rm trans}$) \\
Temperature law  & $400~(r/R_{\rm trans})^{-0.33}$~K\\
Density law & $ 8 \times 10^{14}~(r/R_{\rm trans})^{-0.75}$~H$_2$~mol.~m$^{-3}$ \\
Total gas mass & $\sim 2.1\times10^{-5}$~M$_\odot$ \\
Velocity law & $4 + 12\times \left(\frac{r-R_{\rm trans}}{R_{\rm out} - R_{\rm trans}}\right)$~km~s$^{-1}$\\
Abundances SO, SO$_2$ & $\sim1.5 \times 10^{-6}$, $\sim2.5 \times 10^{-6}$\\
\end{tabular}
\end{table}

While the optical depths in the $^{13}$CO~$J=3-2$ line reach values of $\sim 1$ in our models, the SO$J_K= 8_8-7_7$ emission reaches higher values.
Especially in the innermost regions ($r \leq R_{\rm trans}$), strong optical depth effects are expected to affect these lines. The SO$_2$ lines have average optical depths
peaking at $\sim 0.5$, which indicates some effects from opacity. Nonetheless, these transitions are not strongly affected by optical depth effects and our model confirms that they
constitute a good probe of the excitation temperature and SO$_2$ column density in the lobes.

\section{Discussion}
\label{sec:disc}

\subsection{Gas density and temperature in the extended atmosphere}

\begin{figure*}[th!]
        \centering
        \includegraphics[width=0.9\textwidth]{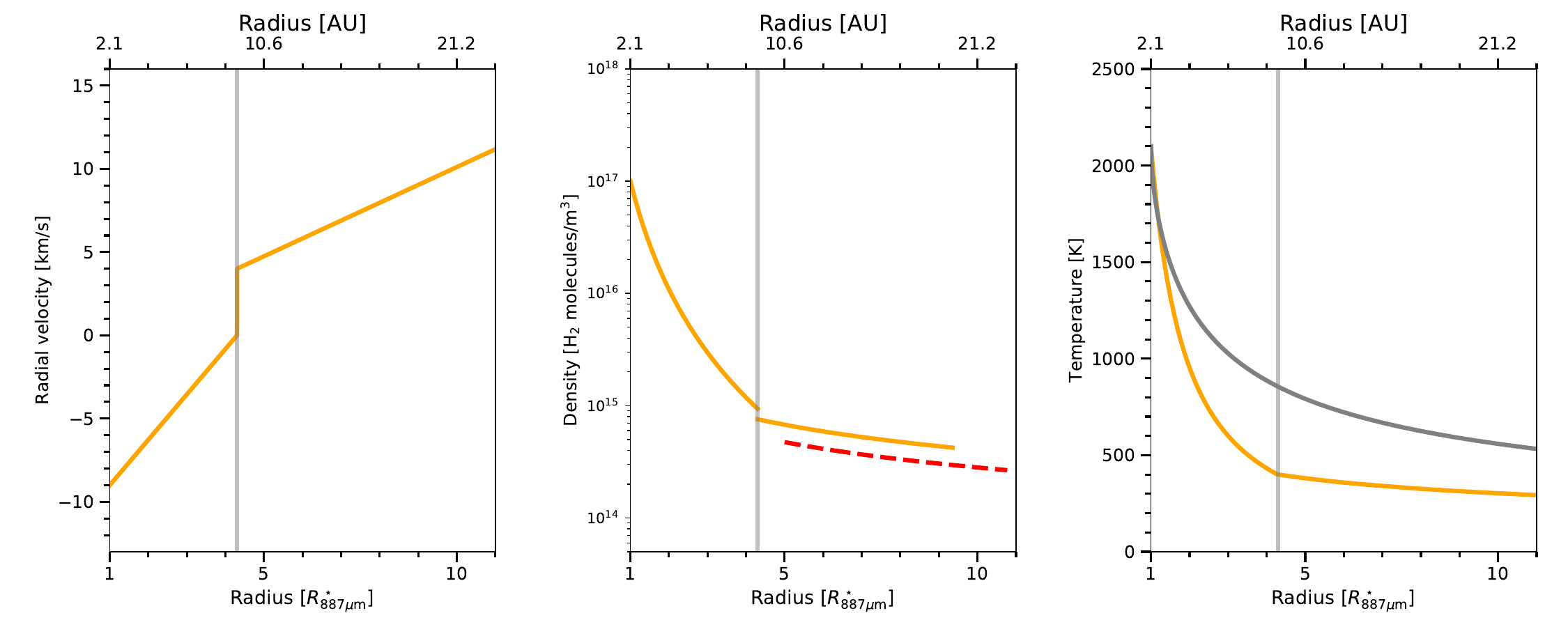}
        \caption{Physical parameters of the radiative transfer model (yellow line) summarized in Table~\ref{tab:model} and with fits to the
        observations shown in Fig.~\ref{fig:lineFits}. The dashed red line shows the
        lower density in the NE lobe. The gray curve in the temperature panel shows a gray atmosphere temperature profile.
        The vertical light gray lines mark $R_{\rm trans}$.}
        \label{fig:model}
    \end{figure*}

The model described in Sect.~\ref{sec:modelRes} and shown in Fig.~\ref{fig:model} was tailored to reproduce the low-excitation lines.
This requires making a compromise of not reproducing the strength of the absorption observed in the CO~$v=1, J=3-2$ line.
As discussed by \cite{Khouri2018},
the relatively strong absorption feature in the CO~$v=1, J=3-2$ line is an indication of this line tracing gas that is cool compared to the stellar continuum temperature. However,
if we decrease the gas temperature in the innermost regions of the model to try to
reproduce this feature, the absorption in the lower-excitation lines $^{13}$CO~$J=3-2$ and SO~$J_K=8_8-7_7$ becomes stronger.
Therefore, we are not able to reproduce the strength of the absorption in
all lines simultaneously, and we conclude that there are aspects of the physical properties of the gas in
the extended atmosphere that are not captured by our model. We prioritize reproducing the low-excitation transitions, which are used to constrain
the properties of gas in the lobes.

One possible way to explain this problem is a more complex distribution of gas with cool and warm components existing at the same distance from the star,
which is different from our model in which the temperature is a function of only the radial distance.
Such a structure is not unexpected and could correspond to gas in updrafts and downdrafts, or an overall clumpy structure, with gas having different properties at the same radius such as seen
in three-dimensional hydrodynamical models \citep{Freytag2023}.
Another possible explanation is that our model does not include the proper stellar radiation field to account for the non-LTE excitation of CO to
excited vibrational states. Our approach for including the star in our model has the shortcoming that the stellar brightness temperature corresponds to the one observed in the continuum at 338~GHz (2100~K) and is the same at all frequencies. In reality, the
radiation from deeper, hotter regions of the star probably contribute significantly to the radiation field at $\sim 2.3$ and $\sim 4.6~\mu$m, which is responsible for exciting CO to higher vibrational states.
Despite this shortcoming, we deem it more important to include a star that provides the correct continuum brightness
temperature at the frequencies of our observations because that has a direct effect on the amount of absorption in all lines.

Both the excitation temperature map and the radiative transfer calculations reveal a steep temperature profile close to the star, with temperatures as low as 600~K reached at $r \sim 3~R_\star$ and 430~K at $r \sim 4~R_\star$.
Such a steep temperature profile
is in agreement with previous results which revealed cold gas close to the Mira~A \citep{Wong2016,Kaminski2017,Khouri2018}, as well as similar results for other oxygen-rich AGB stars, such
as R~Dor and W~Hya \citep{Khouri2024,Vlemmings2017,Ohnaka2025}.
This indicates that low gas temperatures close to the star are a relatively common feature of oxygen-rich AGB envelopes.
For comparison, the radius at which the temperature has dropped to $\sim 430$~K is $\sim 9$ times larger than where it reaches 2100~K in state-of-the-art three-dimensional hydrodynamic models of AGB stars \citep{Freytag2023}.

\subsection{Velocity profile}

We consider the determined velocity profile more uncertain than the density and temperature profiles. There are two main reasons for this. First, the velocity distribution in
the model was only constrained using the line profiles as a function of aperture size, because we did not aim to reproduce the brightness distribution in individual
velocity channels in detail. Second, the radial velocities and the turbulence in the gas both affect the line profiles and
are somewhat degenerate in the models,
with the turbulence also probably varying as a function of radius from the star.
Hence, the exact shape of the velocity
profile should be seen as a representative possibility rather than a full description, but its general characteristics,
such as expanding outer lobes with maximum expansion between $\sim12$~km/s and an in-falling inner shell with average velocities $\sim - 4$~km/s, are trustworthy. 

Additionally, the discontinuity in the profile at $R_{\rm trans}$ does not necessarily represent a region of acceleration of the flow, as in a steady-wind scenario,
but instead of the radial boundary between bodies of gas that experienced different kinematic histories.

\subsection{Redshifted emission peak in the SO$_2$ lines}

In the spectra extracted from small apertures shown in Fig.~\ref{fig:lineFits}, an emission component observed in the SO$_2$ lines at velocities $\sim 58$~km~s$^{-1}$ ($\sim 11$~km~s$^{-1}$ with respect to the systemic velocity
of 47~km~s$^{-1}$)
is not reproduced by the model. The SO line shows emission at similar velocities which might be produced by the same kinematic component of the envelope.
At larger apertures, no additional emission is observed at those velocities except for that already present in the spectrum from the innermost region.

\begin{figure}[h!]
        \centering
        \includegraphics[width=0.48\textwidth]{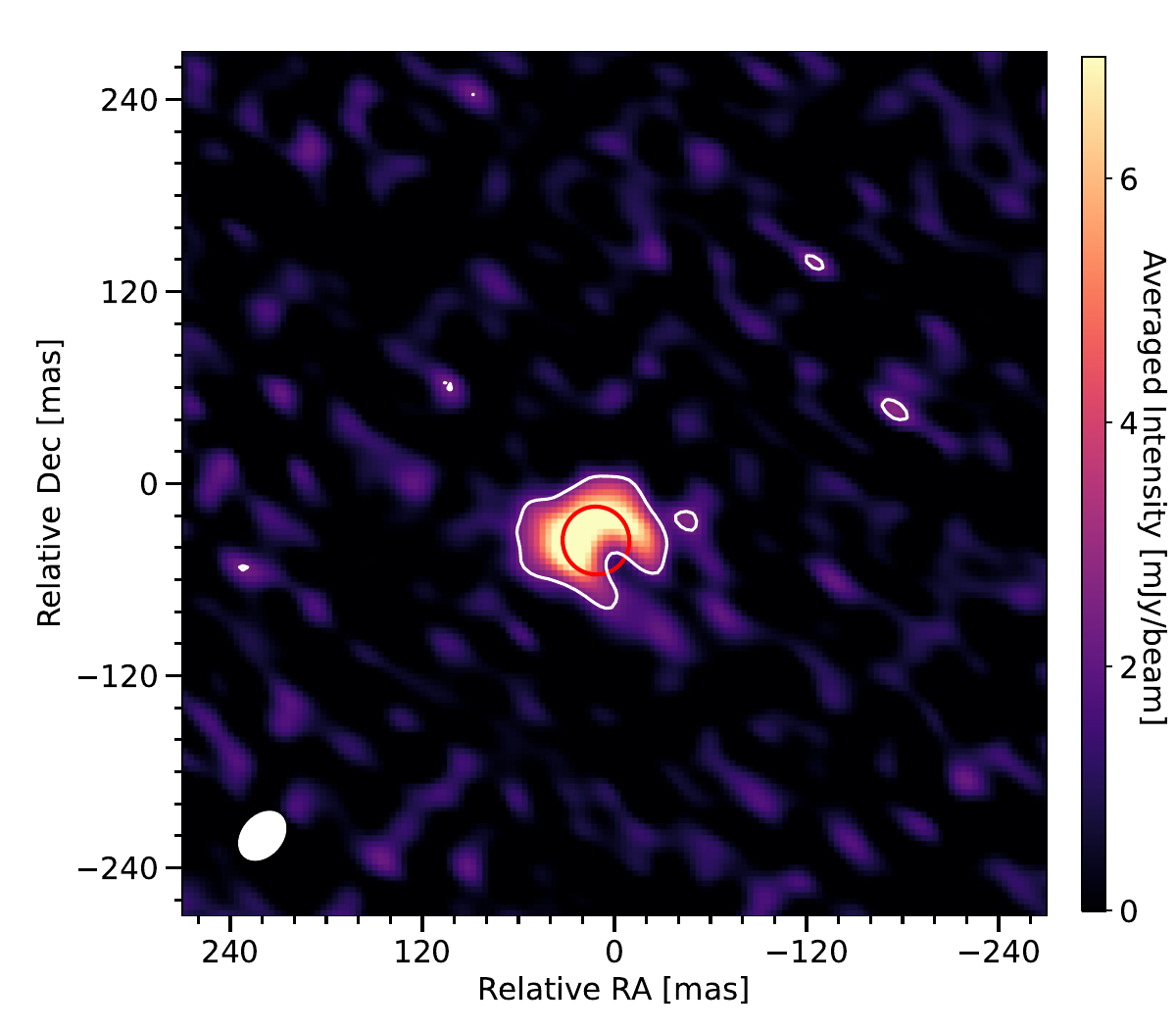}
        \caption{Redshifted emission excess observed in the SO$_2$~$J_{K_{\rm a},K_{\rm c}} = 21_{2,~20}-21_{1,~21}$ line. The red circle represents the stellar disk observed in the continuum emission.
        The white contour marks emission above three times the RMS noise level.}
        \label{fig:SO2_excess}
    \end{figure}

To determine the location
of the gas producing this emission, we averaged the SO$_2$~$J_{K_{\rm a},K_{\rm c}} = 21_{2,~20}-21_{1,~21}$ line over the five channels in which the emission is clearly seen. The resulting map is shown in Fig.~\ref{fig:SO2_excess}.
The emission arises from the direction where the CO~$v=1, J=3-2$ emission also peaks (Fig.~\ref{fig:COv1}).
In contrast to the CO~$v=1, J=3-2$ line, SO$_2$~$J_{K_{\rm a},K_{\rm c}} = 21_{2,~20}-21_{1,~21}$ emission seems to be produced in front of the stellar disk, which is unexpected for cooler gas surrounding a hotter source.
If emission arises from gas in front of the star, the observed redshifted velocity would correspond to gas falling back to the star.
Moreover, an excitation temperature higher than the brightness temperature of the star is expected for thermal emission in this case. However, the molecules responsible for this feature do not seem to have a particularly high excitation
temperature because the feature is strongest in the SO$_2$~$J_{K_{\rm a},K_{\rm c}} = 21_{2,~20}-21_{1,~21}$ line with excitation energy of $\sim 219.5$~K, and its strength decreases in the higher-excitation lines (see Fig.~\ref{fig:lineFits}).
If the emission arises from close to the stellar limb by gas located behind the star, it would be produced by outflowing material.

This feature could also be produced by stimulated emission in principle, if the excitation is non-thermal.
However, it is probably unlikely that maser emission in several SO$_2$ lines at similar velocities
arises from the same region.

\cite{Vlemmings2017} and \cite{Ohnaka2025} also reported emission toward the stellar disk of the AGB star W~Hya and attributed the lines to SO$_2$ transitions.
In the interstellar medium and in protostars, SO and SO$_2$ are known as tracers of shocks \citep[e.g.,][]{PineaudesForets1993,Podio2015}.
A deeper understanding of whether these molecules also trace shocks in the inner regions of the circumstellar envelopes of AGB stars might provide new
tools to studying these crucial innermost regions, where shocks and active chemistry
affect the level populations of molecules. Such an investigation is beyond the scope of this study.

\subsection{Recent mass-loss rate and ejection periodicity}
\label{sec:recentMassLoss}

We find that the ejections observed using ZIMPOL and ALMA must have taken place roughly between 5 and 7 years before the ALMA
observations in November 2017 when considering expansion with deceleration or constant velocity, respectively.
Since the expansion velocity evolution cannot be well constrained based on the observations we report, we are unable to constrain the ejection mechanism.
The observations are consistent with either an energetic event that expelled gas at very high velocities
or the standard dust-driven outflow with relatively fast acceleration to the current expansion velocities.

The density of the gas around the lobes can be estimated by assuming that the excitation temperatures and abundances are similar in the lobes and the gas surrounding them. The
$^{29}$SiO abundance for solar composition is $\sim 1.6 \times 10^{-6}$ without depletion, which is very similar to the value we derive for the SO abundance ($\sim 1.5 \times 10^{-6}$).
Considering LTE excitation at a temperature $\sim 300~K$,
emission in the $^{29}$SiO~$J=8-7$ line is expected to be a factor of ten stronger than in the SO~$J_{K} = 8_8-7_7$ line given the difference in the value of the partition functions,
energy levels, and Einstein-A coefficients of the two transitions.
Therefore, the fact that we see weak emission in the $^{29}$SiO line surrounding the lobes but no emission in the SO line implies
a gas density contrast $\lesssim 10$ under these assumptions.

The fate of the ejected gas is not immediately clear because the material in the lobes might leave the system or not. At a distance of 20~au, the escape velocity from a star with 1~$M_\odot$
is $\sim 9$~km~s$^{-1}$. However, this value can easily be $\sim 12$~km~s$^{-1}$ if the present-day mass of Mira~A is higher ($\sim 1.6~M_\odot$) or if Mira~B is included in the calculation. Hence, the maximum inferred expansion velocities
are tantalizingly close to the expected escape velocity, and the gas might not escape the system without an extra force, such as radiation pressure on dust. In fact, the derived initial velocity for the ballistic trajectory to the NE lobe
($31.6\pm0.6$~km~s$^{-1}$) is within the possible range of escape velocities from the surface of Mira~A, between $\sim 26$ and 35~km~s$^{-1}$ for a present-day mass between 0.8 and 1.5~$M_\odot$.
This implies that gas following trajectories within the uncertainty would either narrowly escape the gravitational field of Mira~A or not. For
a 1.2~$M_\odot$ star (with trajectories shown in Fig.~\ref{fig:expansion}), gas moving ballistically would expand to distances of at least $\sim 70$~au before potentially falling back.
In reality, the problem is complicated by the distribution of previously ejected gas in the circumstellar environment and the effect of Mira~B.
Additionally, even a relatively small amount of radiation pressure on dust would imply that the material indeed leaves the system. However, dust does not seem to be very abundant in the lobes, apart from in the edges,
which puts to question whether the observed lobes are mostly driven by radiation pressure on dust, or created
by an alternative ejection mechanism \citep[e.g.,][]{Judge1991} with radiation pressure on dust playing only a supporting role. Observing the future evolution of the lobes should provide important clues to answering these questions.

It is possible to set constraints on the potential periodicity of ejections of this magnitude. In the case that all gas in the lobes leaves the system, events such as this would contribute to the mass lost by Mira~A.
However, determining the mass-loss rate and history of Mira~A on larger scales is difficult because of the complexity introduced by the gravitational pull, radiation field, and high-velocity outflow of Mira~B \citep{Ramstedt2014}.
Nonetheless, analyses of CO lines indicate a relatively low value for the mass-loss rate of Mira~A. Scaling values from the literature using the distance and CO abundance considered by us, we obtain $1.1\times 10^{-7}$,
$1.5\times 10^{-7}$, $1.9\times 10^{-7}$, $4\times 10^{-7}~M_\odot$~yr$^{-1}$ from values available in \cite{DeBeck2010}, \cite{Planesas1990}, \cite{Ryde2001}, and \cite{Knapp1998}, respectively.
We note that the models from \cite{Ryde2001} and \cite{Planesas1990} have fairly low values for the expansion velocity (2.5 and 3.0~km~s$^{-1}$, respectively),
which are at odds with the observed expansion velocity of the lobes.
The results from \cite{DeBeck2010} and \cite{Knapp1998} imply higher maximum expansion velocities (8.1 and 6.7~km~s$^{-1}$, respectively), which are more comparable to the values we infer for the lobes.
In \cite{Knapp1998}, the outflow of Mira~A is divided into two components, one faster (6.7~km~s$^{-1}$) and the other slower, expanding at 2.4~km~s$^{-1}$ and accounting $\sim 18\%$ of the total rate of mass loss.

On even larger scales, a tail seen in far-UV light likely produced by emission from collisionally excited H$_2$ \citep{Martin2007} is seen to extend $\sim 4$~pc away from the system.
Using the proper motion of the binary pair, \cite{Martin2007} derived that this tail corresponds to mass lost in the last $3\times10^4$~years. However,
hydrodynamical modeling revealed that the ejected gas is decelerated more slowly by the interstellar medium
and that the tail might have been created over $4.5 \times 10^5$~years instead \citep{Wareing2007}. Both works considered a mass-loss rate of $3\times10^{-7}~M_\odot~{\rm yr}^{-1}$ (or $2.6\times10^{-7}~M_\odot~{\rm yr}^{-1}$
at 100~pc) and
an expansion velocity of 5~km~s$^{-1}$. Emission from atomic hydrogen was also detected from part of the tail and the observed
deceleration of the gas implies timescales of $1.2 \times 10^5$~years for its creation \citep{Matthews2008}. The observed UV emission can be reconciled with a constant mass-loss
rate of the order estimated from CO lines \citep{Martin2007,Wareing2007}, but the relatively weak HI emission (accounting for $\sim 20\%$ of the expected amount of gas) raises some questions
about the properties of the gas in the tail.
Based on the study of the tail, the mass-loss rate seems to have remained mostly constant over the last $\sim 10^5$ years. While a mass-loss rate of a few times $10^{-7}~M_\odot~{\rm yr}^{-1}$ seems consistent with the observations,
the analysis and modeling of the tail do not seem to exclude different values of the mass-loss rate and expansion velocity.

Given the complexity of the distribution of gas in the system \citep{Ramstedt2014}, mass-loss rates derived for Mira~A based on emission lines observed with single-dish telescopes are most likely not very accurate. 
Nonetheless, the large-scale rate of mass loss appears to be between $1\times 10^{-7}$ and $4\times 10^{-7}~M_\odot~{\rm yr}^{-1}$, which is not at odds with the mass-loss history
on timescales as long as 10$^5$~years. Hence, we use this range in the following calculations.

If the type of ejection that produced the lobes is not periodic, or if it happens so rarely that it does not contribute significantly to the gas in the system, observing this
event so close to its inception would be very unlikely. Therefore, the scenario preferred by us is that events like this must be relatively common.
Regarding the fate of the ejected material in the lobes, we explore two different scenarios: 1) most of the mass in the lobes will be ejected from the system, 2) most of the mass currently in the lobes will
remain gravitationally bound to the system.

\subsubsection{Scenario 1: Periodic ejections that contribute significantly to the total mass-loss rate}
\label{sec:scenario1}

The gas mass traced by molecular lines in the lobe ($\sim 2.1 \times 10^{-5}~M_\odot$)
can account for the vast majority, or even all, of the between $1\times 10^{-7}$ and $4\times 10^{-7}~M_\odot~{\rm yr}^{-1}$ estimated to be lost by $o$~Ceti, if this type of ejection occurs periodically.

In order to account for all the mass lost by Mira~A, a timescale between 50 and 200~years would be required for ejections of this type and magnitude. If we consider the two lobes to have been ejected by different events,
the timescale would be reduced by half. However, it would be unlikely that two lobes would be created within roughly one year of each other
if this was a stochastic process with a characteristic timescale $\gtrsim 25$~years.
Hence, given the coincidence of the estimated ejection times for the two lobes, the most logical conclusion at this point is that the lobes were created
by the same event. According to our estimates, this constituted a short period of enhanced mass loss
that started some time between 2010 and 2012 and could have lasted for a few years.
An inspection of the light curve of the Mira system between 2005 and nowadays using the AAVSO international database reveals that Mira displayed maxima and minima brighter by $\sim 1$~mag in visible light
between late 2010 and early 2012 (over $\sim 2.5$ pulsation cycles). Whether this is associated with the observed ejection or not is unknown. A similar pattern in the light curve can be seen between late 2019 and late 2021. No associated
enhancement in mass-loss rate has been identified so far, but observations of the system in the coming years will reveal if that was the case. Over the last $60$ pulsation cycles (55~years), $\sim 20\%$ of the maxima are brighter by
a comparable amount.

Given the periodicity we estimate, it is unlikely that we would witness another an ejection of this magnitude in the near future. However, this estimate is inversely proportional to the uncertain mass-loss rate on larger scales. To improve our understanding of this type of ejection process and its effect on the circumstellar envelope and the mass-loss rate of Mira A, a better description of the physical parameters of the complex circumstellar environment is necessary. If the mass-loss rate has been underestimated or if less massive ejections happen more frequently, observations of the system in the coming decades should reveal additional ejections.

Regarding past events,
an expansion velocity of $\sim 11$~km~s$^{-1}$ and a timescale between 50 and 200 years corresponds to radial distance between 115~au (1\farcs15) and
460~au (4\farcs6), in radius for a potential previous ejection.
Hence, the signature of these periodic ejections could in principle be found in the envelope with current instruments. However, the strong effect Mira~B has on the 
distribution of the gas \citep[e.g.,][]{Ramstedt2014} and
the orbital period of the system of $\sim 500$~years \citep{Prieur2002}, make it difficult to identify gas ejected by previous events
in the larger-scale circumstellar envelope of Mira. Nonetheless, \cite{Wong2016} reported a structure observed in the SiO~$J=5-4$ line to the southwest of Mira~A and located at a projected distance of $\sim2\farcs7$ consistent
with our estimate for a possible previous ejection.
This structure could have been produced by the same type of ejection process reported by us, but this is highly speculative at this point.

\subsubsection{Scenario 2: An ejection that does not escape the system}

If ejections of this type are relatively common but the gas does not leave the system, the periodicity estimated based on the mass-loss rate is not meaningful. 
Moreover, accretion of the ejected material by Mira~A or Mira~B would be required in this case. The relatively low accretion rate of Mira~B \citep[$\sim 10^{-10}~{\rm M}_\odot~{\rm yr}^{-1}$,][]{Sokoloski2010} implies
that most of the material would have to be re-accreated by Mira~A.
Given the high gas mass we derive in the inner regions of the circumstellar envelope of
Mira~A ($8 \times 10^{-5}~{\rm M}_\odot$), this seems possible. In this case, the ejection timescale would be difficult to directly constrain, but would likely be relatively short ($\lesssim 100$~years)
to make the serendipitous observation of this event likely. A re-accretion rate $> 10^{-7}~{\rm M}_\odot~{\rm yr}^{-1}$ would be expected. Such a gas inflow is not obvious from the data, but it is also
unclear from our analysis whether this possibility can be excluded. However, we deem it unreasonable for gas to be accreted by Mira~A at rates 1000 times higher than by Mira~B, given that material
reaches distances from Mira~A which are larger than the orbital separation.

Based on these considerations, our preferred scenario is that the ejected gas will likely leave the system.
In this case, the mass loss of Mira~A would likely be dominated by relatively rare, massive ejections, contrary to our current expectation for AGB stars of variability on timescales comparable to the pulsation period.
Additionally, it is worthwhile noting that an X-ray flare was observed toward Mira~A in late 2003 \citep{Karovska2005}. This unexpected occurrence could in principle be associated with mass ejection.
However, given the time delay between the detected flare and the ejection times estimated by us, we believe a direct connection between this specific X-ray flare and the creation of the lobes unlikely. 
It is not known how often Mira~A produces similar X-ray flares.

\subsection{Molecular abundances}
\subsubsection{SO and SO$_2$}

The combined abundances of SO and SO$_2$ ($\sim 4 \times10^{-6}$) correspond to about seven times lower than that of $^{13}$CO.
For solar composition, the sulfur abundance with respect to H$_2$, $\sim 2.6 \times 10^{-5}$, implies a $^{13}$C/S of only $\sim 1.2$ for $^{12}{\rm C/^{13} C}=13$ \citep{Khouri2018}.
Hence, the total abundance derived by us accounts for a relatively small fraction ($\lesssim 20\%$) of the sulfur expected for solar composition.

The SO and SO$_2$ abundances cannot be high enough to account for the majority of sulfur in the envelope because that would lead to very strong absorption and emission, as shown in Fig.~\ref{fig:SO+SO2_abundances}.
However, given the complex morphology, the abundance can be high in morphological components that do not contribute to absorption against the star, such as the lobes. We also find that the abundances in
the lobes need to be at least $\sim 10^{-6}$ to produce the observed emission. For the Inner Region, which produces absorption against the star, we find an upper limit for the
abundances of $\sim 3 \times 10^{-7}$.

Observations of SO and SO$_2$ in oxygen-rich AGB stars reveal differences between stars with high and low mass-loss rates. On the one hand, the combined abundances of these two molecules
seem to reach stable levels close to the star and account for all the available sulfur in sources with low mass-loss rates ($\sim 10^{-7}$~M$_\odot$~yr$^{-1}$). On the other hand,
the SO$_2$ abundance peaks at several hundred stellar radii from the central source and the two molecules do not account for the expected amount of sulfur in stars with higher mass-loss rates
 \citep[$\sim 10^{-6}$~M$_\odot$~yr$^{-1}$,][]{Danilovich2016}. The classification of Mira in this context is not clear. Despite the average mass-loss rate on longer timescales being relatively low,
 the gas densities in the lobes are as high as expected in a source
 with high mass-loss rate. The observed abundances account for only a fraction of the expected available
 sulfur, in agreement with what is found for sources with higher mass-loss rates. While the lower abundances in the inner few stellar radii are in agreement with the low values found by \cite{Danilovich2016}
 for SO in R~Cas and IK~Tau (with $\dot{M} \gtrsim 10^{-6}$~M$_\odot$~yr$^{-1}$), we find that abundances of the order of 10$^{-6}$
 are reached much closer to Mira~A ($\sim 10$~AU) than in those sources (at tens or hundreds of AU).
The abundances we find are significantly higher than expected from chemical equilibrium models, with expected
peak values $\sim 10^{-7}$ and $\sim 10^{-9}$ between 2 and 5~$R_\star$ for SO and SO$_2$, respectively \citep{Agundez2020}.
More complex chemical model calculations \citep{VandeSande2022,VandeSande2023} reproduce these higher abundances, but do not show lower abundances close to the star
for any mass-loss rates unless the radiation field of a hotter ($T \gtrsim 4000$~K) companion is also added.
We also note that these models do not consider any companions at the separation of Mira B.
It is unclear if activity from Mira~A can mimic these results.

\subsubsection{AlO, AlF, and PO}

Chemical equilibrium models \citep{Agundez2020} predict
the AlF abundance in an oxygen-rich AGB star to reach a peak of $\sim 1 \times 10^{-7}$ between 3 and 10~$R_\star$.
For AlO, the maximum abundance is found to be $\sim 1 \times 10^{-8}$ between 1 and 3~$R_\star$, with a rather steep fall to a value of $\sim 1\times10^{-18}$ at 10~$R_\star$. Finally,
for PO the abundance is found to peak at $\sim 4 \times 10^{-7}$ between 2 and 5~$R_\star$, and to fall to $\sim 1\times10^{-10}$ at 10~$R_\star$.

The abundances we derive for AlF and AlO reach values about two order of magnitude below the expected peak values. This could be caused by depletion of Al into dust grains.
Interestingly, the derived abundance distribution of AlO (Fig.~\ref{fig:AlO}) shows no sign of the precipitous decline predicted by the models, with the abundance peak located at $\sim 7~R_\star$.
For PO, the peak abundance we find is in agreement with the value expected from chemical equilibrium models, but we are unable to confirm the expected decline in abundance at $\sim 5~R_\star$ because
the relatively low signal-to-noise of the observed emission prevents us from inferring whether the PO abundance remains high farther out than $\sim 7~R_\star$.

We note that the gas temperatures adopted in the models of \cite{Agundez2020} are significantly higher than what we find, with $\sim 1000$~K being reached at 5~$R^{\rm IR}_\star$ in comparison to
$\sim 500$~K at $3.5~R_\star$ in the profile we find. In this comparison, we assume a conversion between the stellar radius in the IR ($R^{\rm IR}_\star$) and the sub-mm ($R_\star$) of $\sim 1.4$ based on
the stellar radii discussed in \cite{Vlemmings2019}.

\subsection{Polarized light images}

The dramatic change in the relative brightness of the structures identified in the polarized-light images as a function of time (Fig.~\ref{fig:pol_Light}) is puzzling.
Dust to the south of the star is seen to produce strong polarized light in October 2016 and October 2018, while it is weakly detected in 2017 and
strongly outshined by the edge of the NE lobe (and barely detected overall) in January 2021.
The brightness of the edge of the NE lobe that becomes extreme in 2021, while it is relatively low in 2018 and 2023 is also puzzling.
Explaining this type of variation with changes in the grain sizes or properties would require somewhat abrupt changes that are not expected at
relatively large distances ($r > 10$~au) from the central star.

To explain these observations we consider changes in the orientation of the structures, which would cause the scattering angle of the light detected by us to vary as a function of time.
However, given that the expansion on the plane of the sky is so well described by a constant projected velocity,
it would be difficult to imagine a scenario in which the structures rotate significantly with
respect to the line of sight while maintaining the appearance of constant expansion on the plane of the sky.

Considering a radially expanding outflow,
the expanding structures would have scattering angles with respect to us constant in time. Moreover, the scattering phase function and the polarization efficiency for that given scattering angle should not
vary if the size and composition of the grains are not changing significantly.
Even if we assume that the grain properties are changing as a function of time, the observed changes in polarized light remain difficult to explain.
For grains that are smaller than about the wavelength of observations divided by 2$\pi$
(corresponding to grains of 0.1~$\mu$m in this case), scattering is roughly equally likely in all directions and the scattered radiation has maximum polarization for scattering angles of $\sim 90^\circ$. For larger grains,
both the scattering phase function and the polarization efficiency have more complex behaviors with scattering angle.
 The dramatic changes in brightness of the polarized light would require large grains ($> 0.3~\mu$m) that change sizes significantly over two years. 
 At $r > 10$~au from the central star, such rapid dust growth is no longer expected. For instance, grain size stabilizes at $\sim 2.5~R_\star$ in the wind-driving models presented by \cite{Hoefner2022}.
 
We speculate that the most obvious explanation for the observed variation is a lighthouse
effect, in which the stellar radiation field is not spherically symmetric, and structures that receive more direct stellar light at a given time are seen to shine brighter in polarized light. Shocks happening close to the star and non-uniform distribution
of dust grains or gas
close to the star that block stellar light are potential mechanisms to create an asymmetric stellar radiation field, but whether these effects are sufficient to produce the observed effects is unclear and should be investigated in future studies.

How such a lighthouse effect would affect the very periodic light curve of Mira~A at visible wavelengths is unclear, but it could be related to the brightness variations between cycles discussed in Sect.~\ref{sec:scenario1}.
Given the cadence of the observations we report, between one and two years,
and the relatively small number of epochs, it is not possible to assess whether there is any periodicity to the brightening of the different regions in polarized light.
Monitoring the polarization degree around Mira~A will allow this to be clarified.

\section{Conclusions}
\label{sec:conc}

We reported observations of light polarized by dust scattering and of molecular line emission that reveal two expanding lobes close to Mira~A. The observed expansion
is consistent with the outflow having either a constant velocity or a decelerating profile. These two scenarios would correspond to the standard dust-driven paradigm for AGB outflows
or to an explosive ejection, respectively. We constrained the ejection time to between 2010 and 2012, but we were not able to determine the ejection mechanism. 

Our analysis of the molecular lines reveals high gas masses in the extended atmosphere ($8 \times 10^{-5}~{\rm M}_\odot$) and in the lobes ($2 \times 10^{-5}~{\rm M}_\odot$).
Given the low accretion rate of Mira~B and
the potential effect of additional forces, we consider that the most likely scenario is that the ejected material leaves the system.
If ejection events of this magnitude are periodic and indeed contribute to mass loss, the associated timescale would be between 50 and 200~years. In the unlikely case that the gas does not leave the system, it must be
mostly re-accreated by Mira~A. If the mass-loss rate of the system provided in the literature is not accurate or if similar but less massive events are more common, ejections of this type can happen more often.

Our models constrain the abundances of SO and SO$_2$ in the lobes to be $\sim 1.5 \times 10^{-6}$ and
$\sim 2.5 \times 10^{-6}$, while we find abundances of $2\times10^{-10}$, $6.5\times10^{-10}$, and $4\times10^{-7}$ for AlO, AlF, and PO assuming LTE and optically thin emission.
We also find that the SO and SO$_2$ abundance must be lower close to the star, because otherwise stronger emission from the inner regions would be expected.
Compared to predictions from chemical equilibrium models,
the peak abundances we derive are higher by one and three orders of magnitude for SO and SO$_2$, respectively; lower by two orders of magnitude for AlF and AlO; and in agreement for PO.
Calculations including chemical reaction networks reproduce the observed peak abundances of SO and SO$_2$, but would require a hotter companion close to Mira~A to explain the lower abundances in the inner regions.
It is unclear whether activity from Mira~A can be responsible.

The polarized-light images reveal strong variation in brightness of the different features in the six images obtained. The cause for this is unclear.
We put forward the suggestion that an asymmetric stellar radiation
field could be preferentially illuminating specific regions of the circumstellar envelope at a given time in a lighthouse-like effect.
If this is indeed the case, it would have significant consequences for the interpretation of polarized light from this system and other AGB stars similar to Mira~A.

\begin{acknowledgements}
This paper makes use of the following ALMA data: ADS/JAO.ALMA\#2017.1.00191.S, ADS/JAO.ALMA\#2018.1.00749.S, and ADS/JAO.ALMA\#2022.1.01071.S.
ALMA is a partnership of ESO (representing its member states), NSF (USA) and NINS (Japan),
together with NRC (Canada), MOST and ASIAA (Taiwan), and KASI (Republic of Korea), in cooperation with the Republic of Chile. The Joint ALMA Observatory is operated by ESO, AUI/NRAO and NAOJ.
We gratefully acknowledge the contributions of the AAVSO observer community, whose
photometric data and metadata resources were used in this study and made available through the AAVSO’s scientific archives.
T.K. acknowledges support from the Swedish Research Council through grants 2019-03777.
TD is supported in part by the Australian Research Council through a Discovery Early Career Researcher Award (DE230100183).
\end{acknowledgements}

\bibliographystyle{aa}
\bibliography{bibliography_2}

\begin{appendix}
\onecolumn

\section{Comparison of data from 2017 and 2018 }

   \begin{figure*}[h!]
        \centering
        \includegraphics[width=0.6\textwidth]{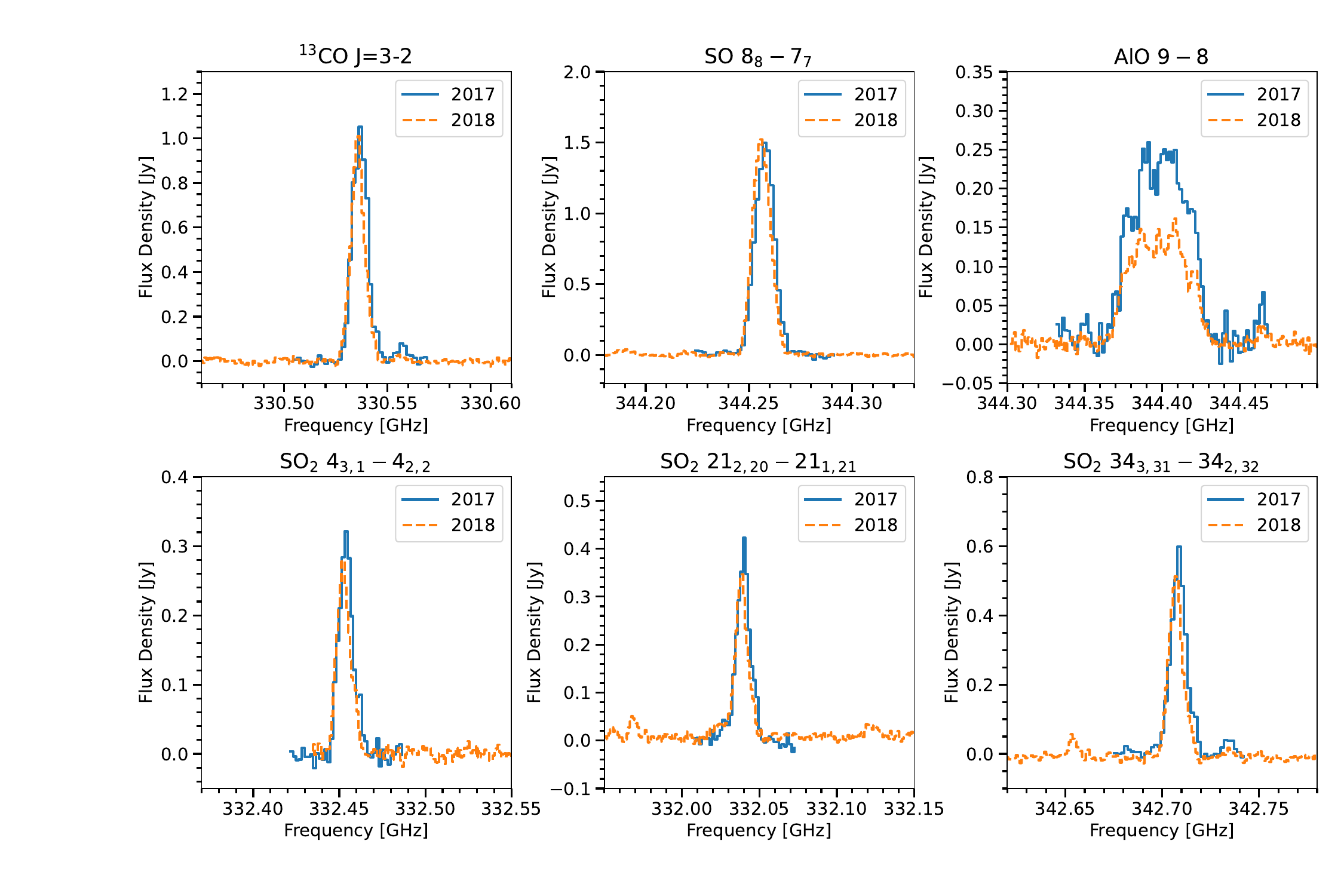}
        \caption{Comparison of observations acquired in 2017 and 2018. The observations from 2017 were convolved to the same spatial resolution as those from 2018
        and the emission was computed using a circular aperture centered on the continuum peak and with diameter equal to $0\farcs4$.}
        \label{fig:resOutFlux}%
    \end{figure*}


\section{SO and SO$_2$ abundance profiles}

   \begin{figure*}[h!]
        \centering
        \includegraphics[width=0.55\textwidth]{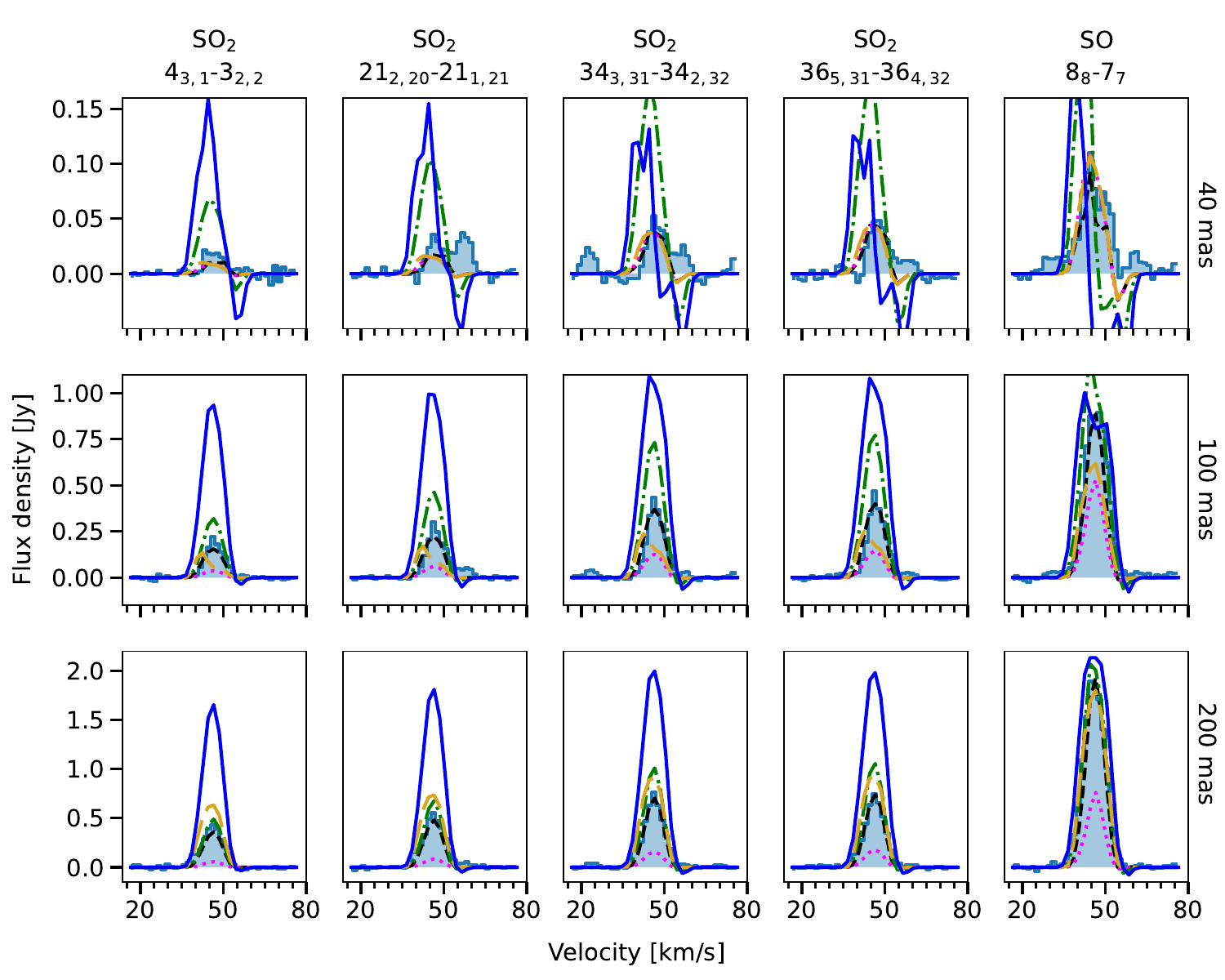}
        \includegraphics[width=0.27\textwidth]{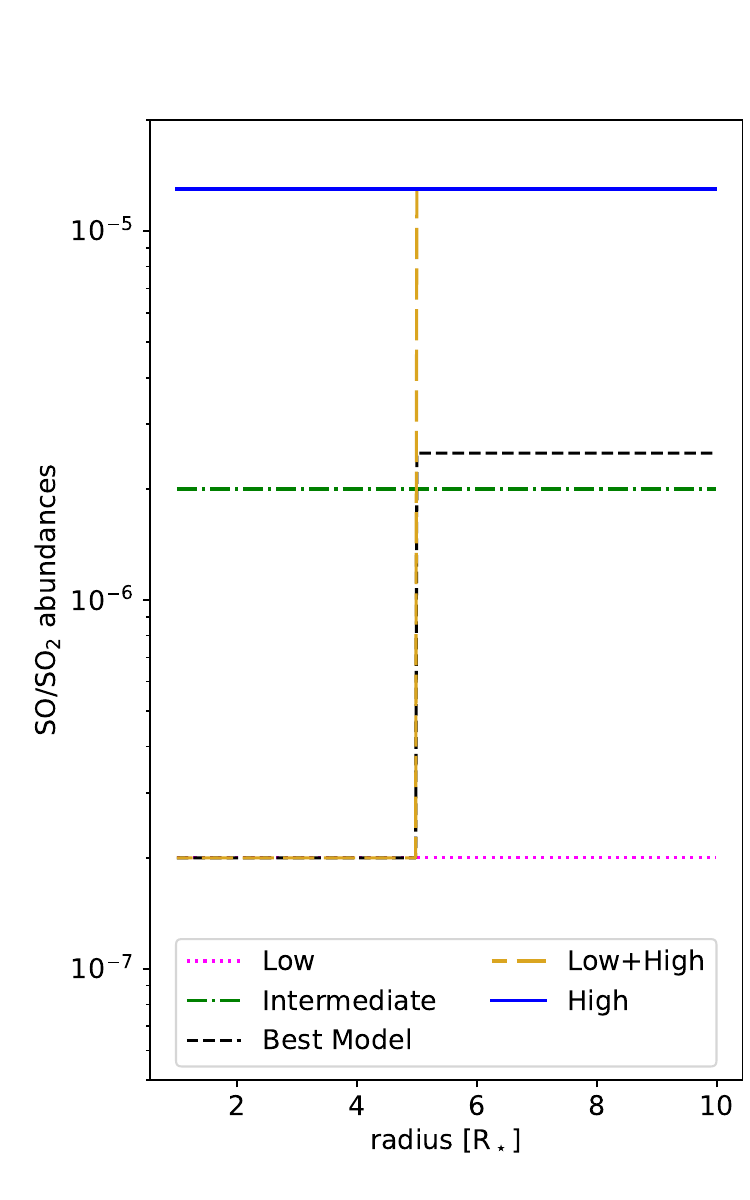}
        \caption{Predicted SO and SO$_2$ line emission (left panels) considering different abundance profiles (right panel) and the density, temperature, and velocity distributions presented in Table~\ref{tab:model}.
        The line emission is shown with the same line color and style as the corresponding abundance profile.}
        \label{fig:SO+SO2_abundances}%
    \end{figure*}


\section{Adopted partition functions and transitions constants}
\label{sec:transParam}
The energy of the upper levels and the Einstein-A coefficients were obtained from the Cologne Database for Molecular Spectroscopy (CDMS) or from the ExoMol database. 
When the partition functions listed in CDMS did not extend to high-enough temperatures ($\sim 1000$~K), we retrieved the partition function and the transition parameters from the ExoMol database.
Care has to be taken that the same levels are taken into account in the partition function and on the calculation of the level populations (for instance, whether hyperfine splitting was accounted for or not) .

For SO and SO$_2$, we retrieved the transition properties and the partition function provided by \cite{Brady2024} and \cite{Underwood2016} through the ExoMol database.

For AlO, the transition properties and the partition function listed in the CDMS were used, with data from \cite{Torring1989,Yamada1990,Goto1994} and \cite{Patrascu2015}.

For PO, we obtained the transition properties and the partition function provided by \cite{Prajapat2017} through the ExoMol database.

For AlF, we used the transition properties and the partition function listed in the CDMS, with data compiled from \cite{Wyse1970,Hoeft1970,Maki1982,Hedderich1992} and \cite{Yousefi2018}.

\section{Line emission, optical depth, and abundance maps}

   \begin{figure*}[h!]
        \centering
        \includegraphics[width=0.65\textwidth]{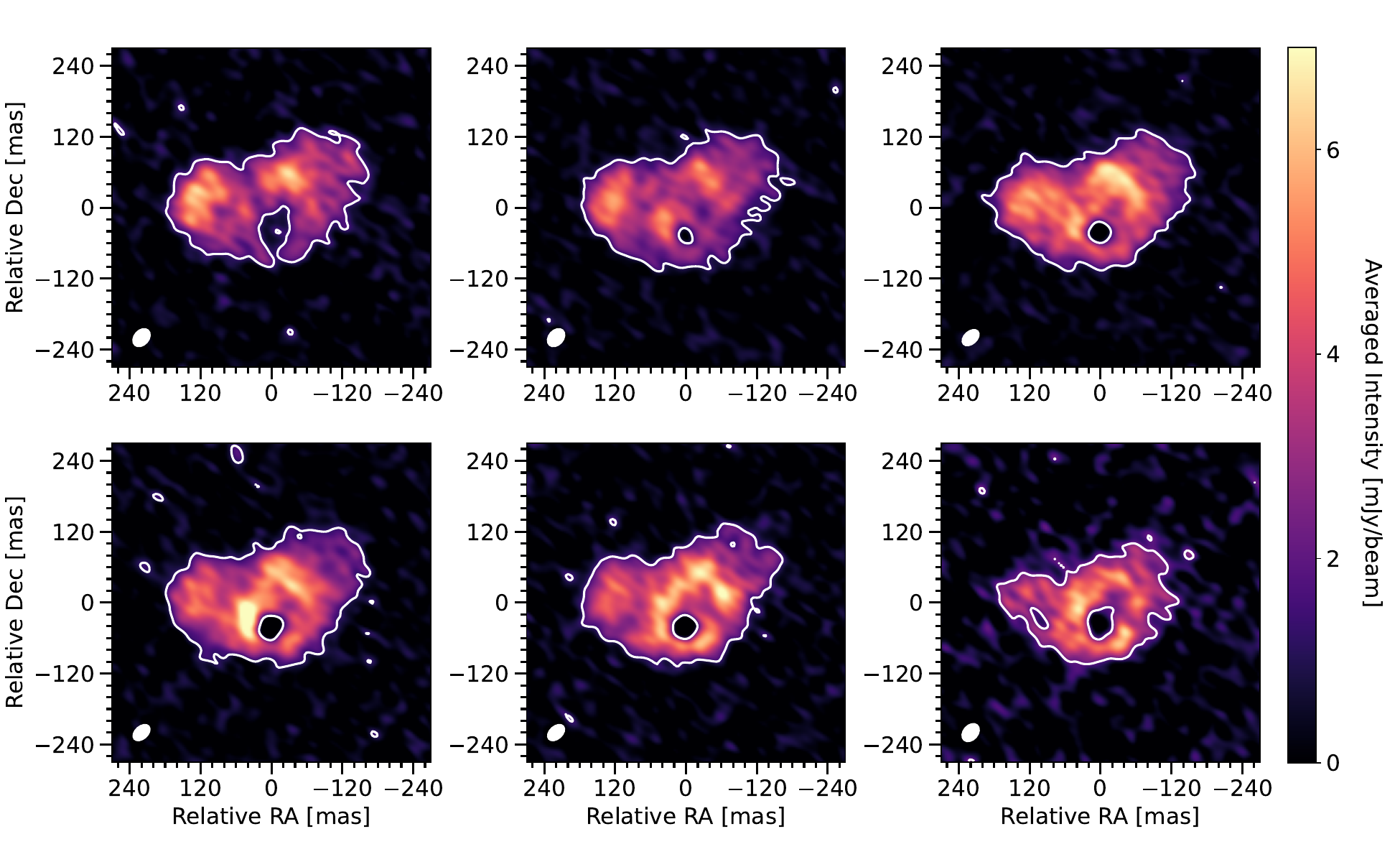}
        \caption{SO$_2$ line emission maps. The panels show lines $J_{K_{\rm a}, K_{\rm c}} = 4_{(3,1)}-3_{(2,2)}$~(top left), $21_{(2,20)}-21_{(1,21)}$~(top middle), $34_{(3,31)}-34{(2,32)}$~(top right),
        $36_{(5,31)}-36_{(4,32)}$~(bottom left), $40_{(4,36)}-40_{(3,37)}$~(bottom middle), and $44_{(5,39)}-44_{(4,40)}$~(bottom right).}
        \label{fig:SO2_maps}%
    \end{figure*}

   \begin{figure*}[h!]
        \centering
        \includegraphics[width=0.65\textwidth]{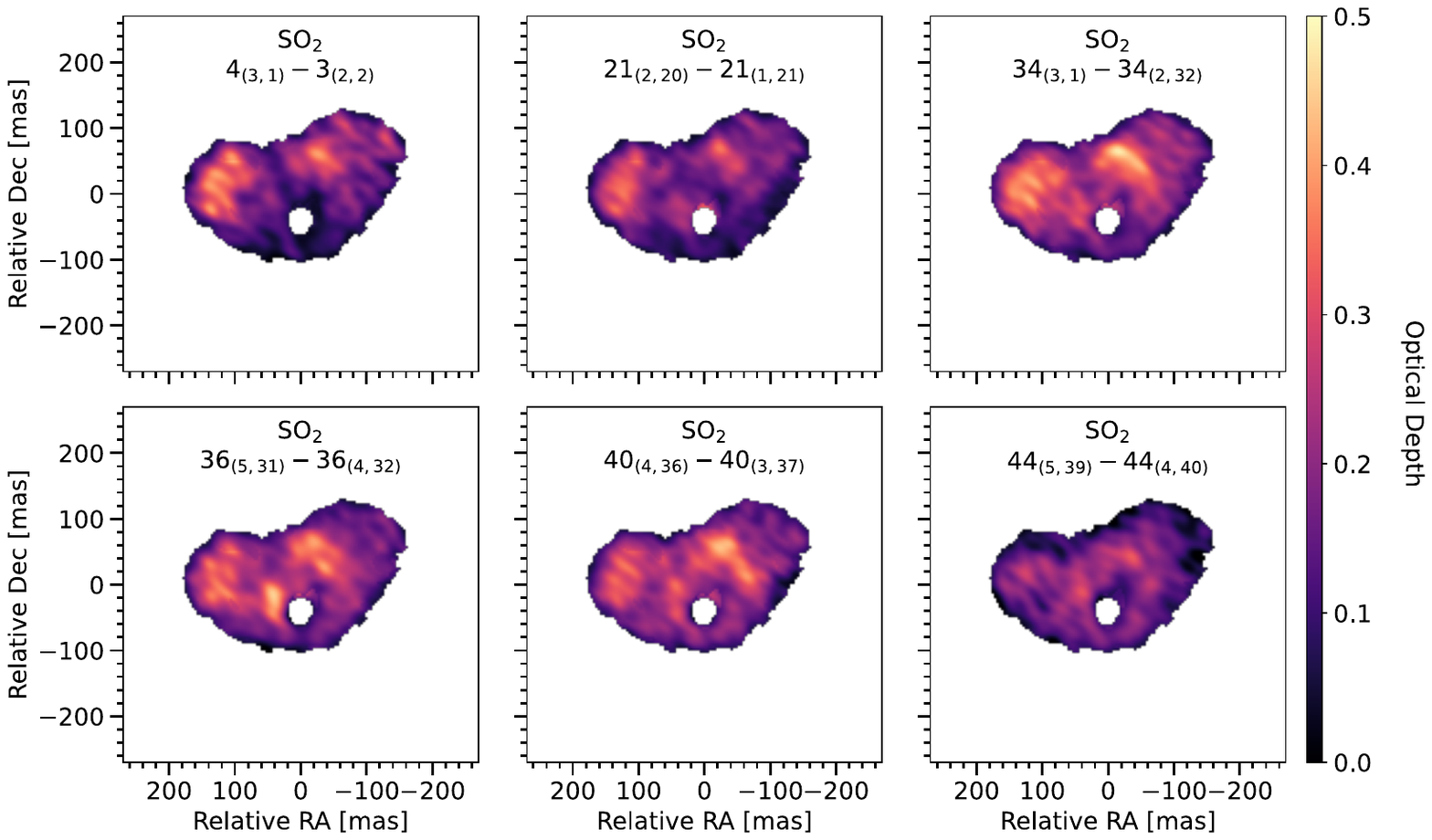}
        \caption{SO$_2$ optical depth maps. The panels show lines $J_{K_{\rm a}, K_{\rm c}} = 4_{(3,1)}-3_{(2,2)}$~(top left), $21_{(2,20)}-21_{(1,21)}$~(top middle), $34_{(3,31)}-34{(2,32)}$~(top right),
        $36_{(5,31)}-36_{(4,32)}$~(bottom left), $40_{(4,36)}-40_{(3,37)}$~(bottom middle), and $44_{(5,39)}-44_{(4,40)}$~(bottom right).}
        \label{fig:SO2_tau}%
    \end{figure*}

   \begin{figure*}[h!]
        \centering
        \includegraphics[width=0.9\textwidth]{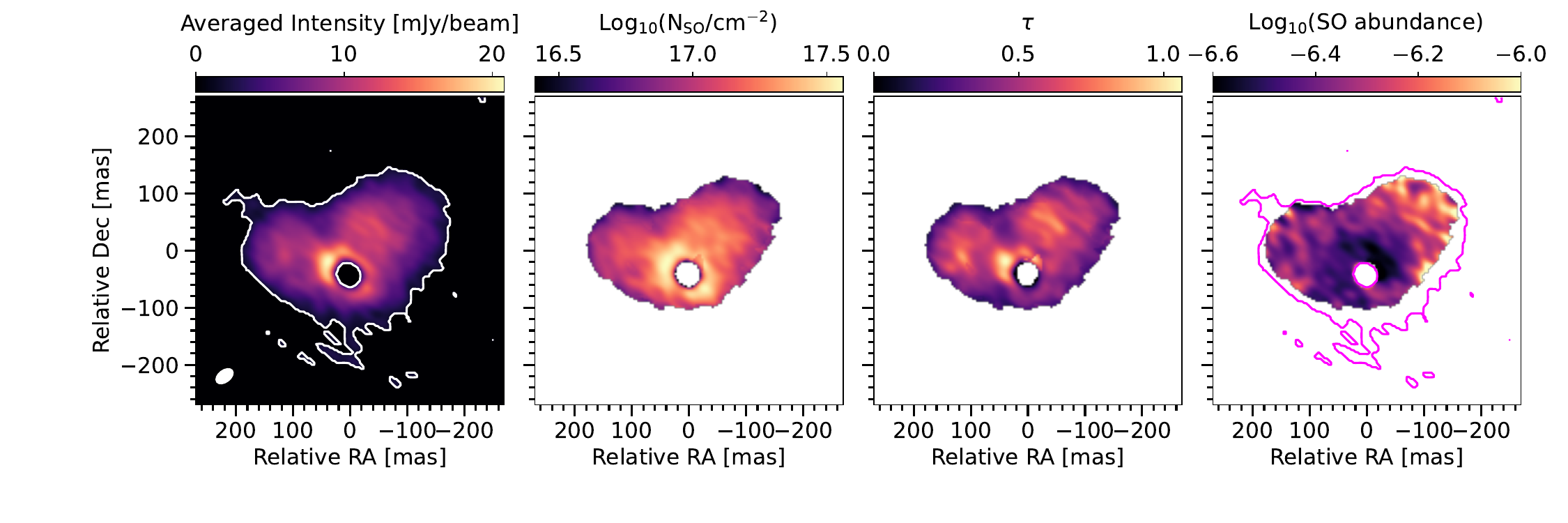}
        \caption{Maps of SO $J_K = 8_8 - 7_7 $ line emission (left), column density (middle left), averaged line optical depth (middle right), and abundance (right).}
        \label{fig:SO_tau}%
    \end{figure*}

   \begin{figure*}[h!]
        \centering
        \includegraphics[width=0.9\textwidth]{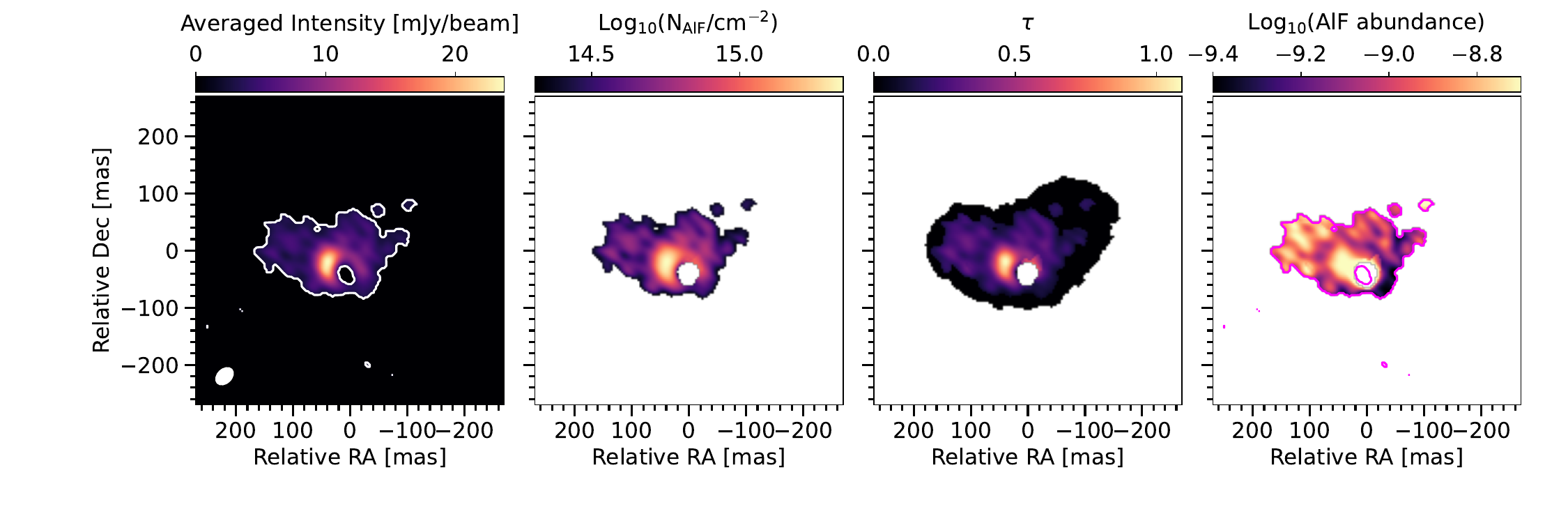}
        \caption{Maps of AlF $J=10-9$ line emission (left), column density (middle left), averaged line optical depth (middle right), and abundance (right).}
        \label{fig:AlF}%
    \end{figure*}

   \begin{figure*}[h!]
        \centering
        \includegraphics[width=0.9\textwidth]{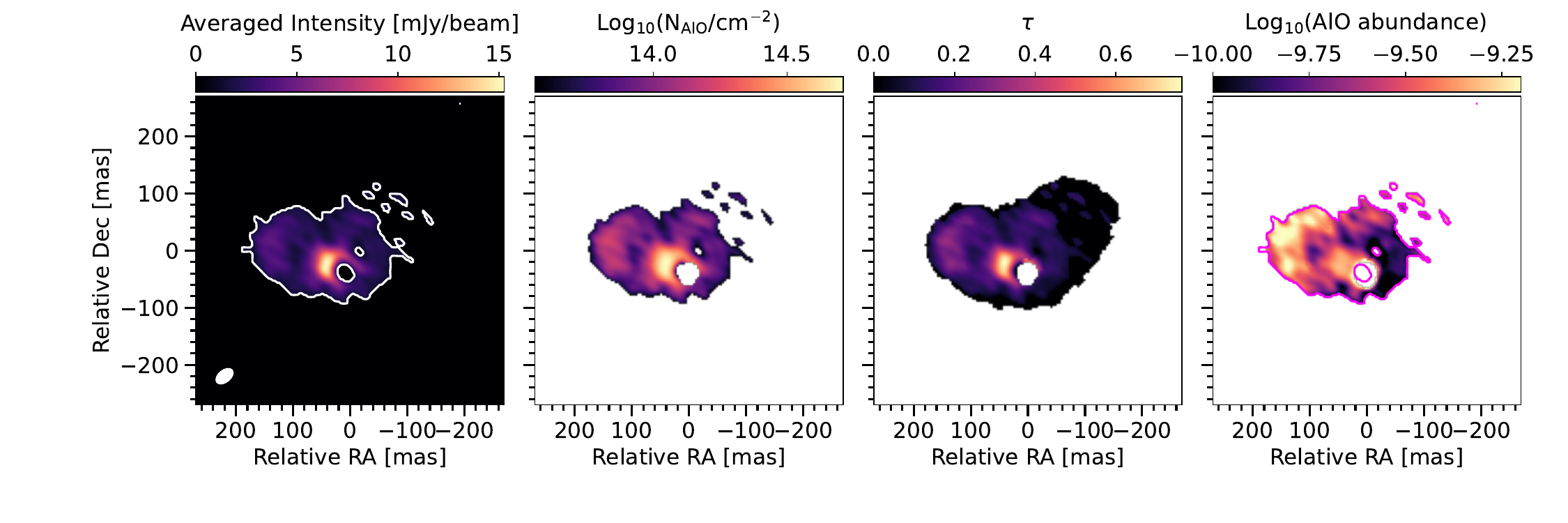}
        \caption{Maps of AlO $N=9-8$ line emission (left), column density (middle left), averaged line optical depth (middle right), and abundance (right).}
        \label{fig:AlO}%
    \end{figure*}

   \begin{figure*}[h!]
        \centering
        \includegraphics[width=0.9\textwidth]{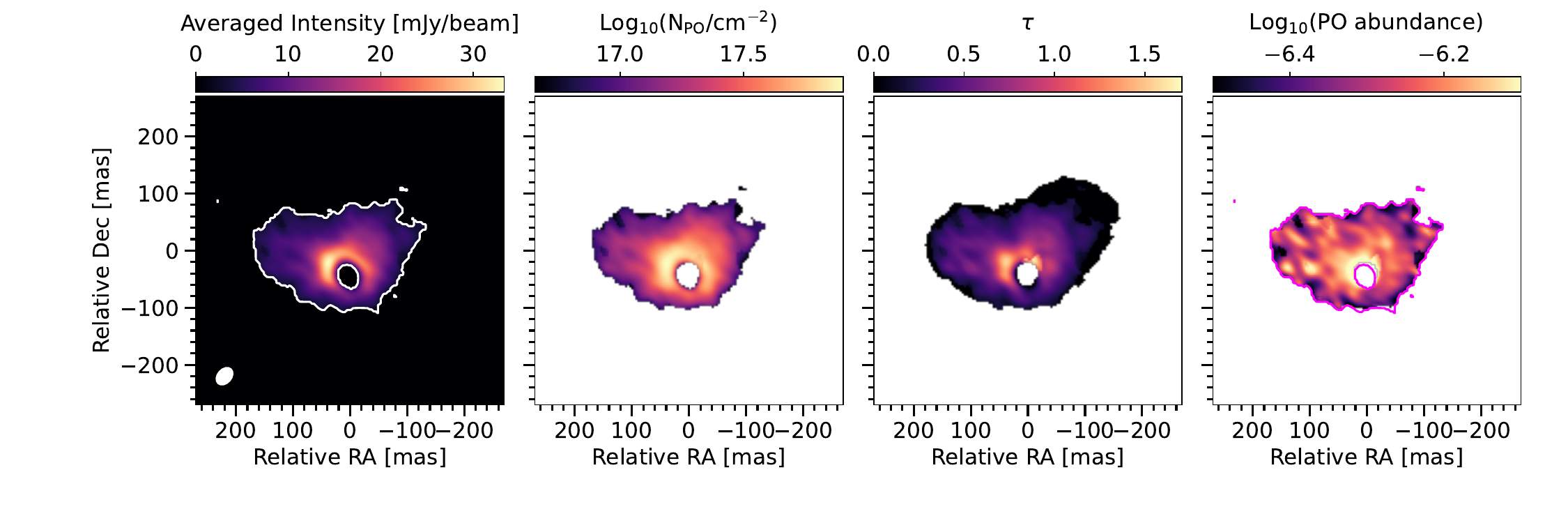}
        \caption{Maps of PO $^2\Pi_{3/2}~J, F= 8,8 - 7,7$ line emission (left), column density (middle left), averaged line optical depth (middle right), and abundance (right).}
        \label{fig:PO}%
    \end{figure*}

   \begin{figure*}[h!]
        \centering
        \includegraphics[width=0.85\textwidth]{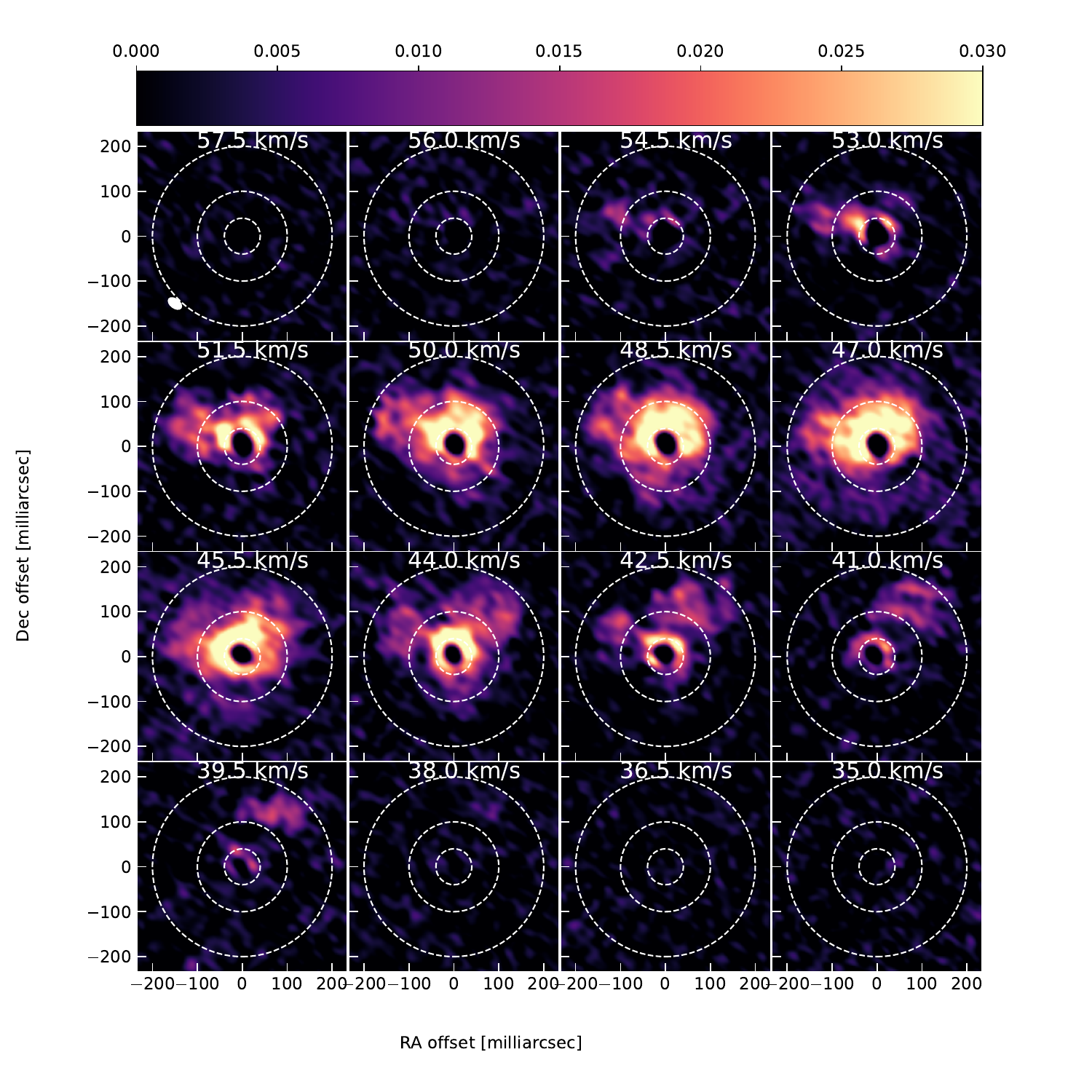}
        \caption{Channel maps of $^{13}$CO~$J=3-2$ emission. The white dashed circles with radii 40, 100, and 200~mas indicate the region from where spectra were extracted to be compared to the radiative transfer models.}
        \label{fig:13CO-channels}%
    \end{figure*}

   \begin{figure*}[h!]
        \centering
        \includegraphics[width=0.85\textwidth]{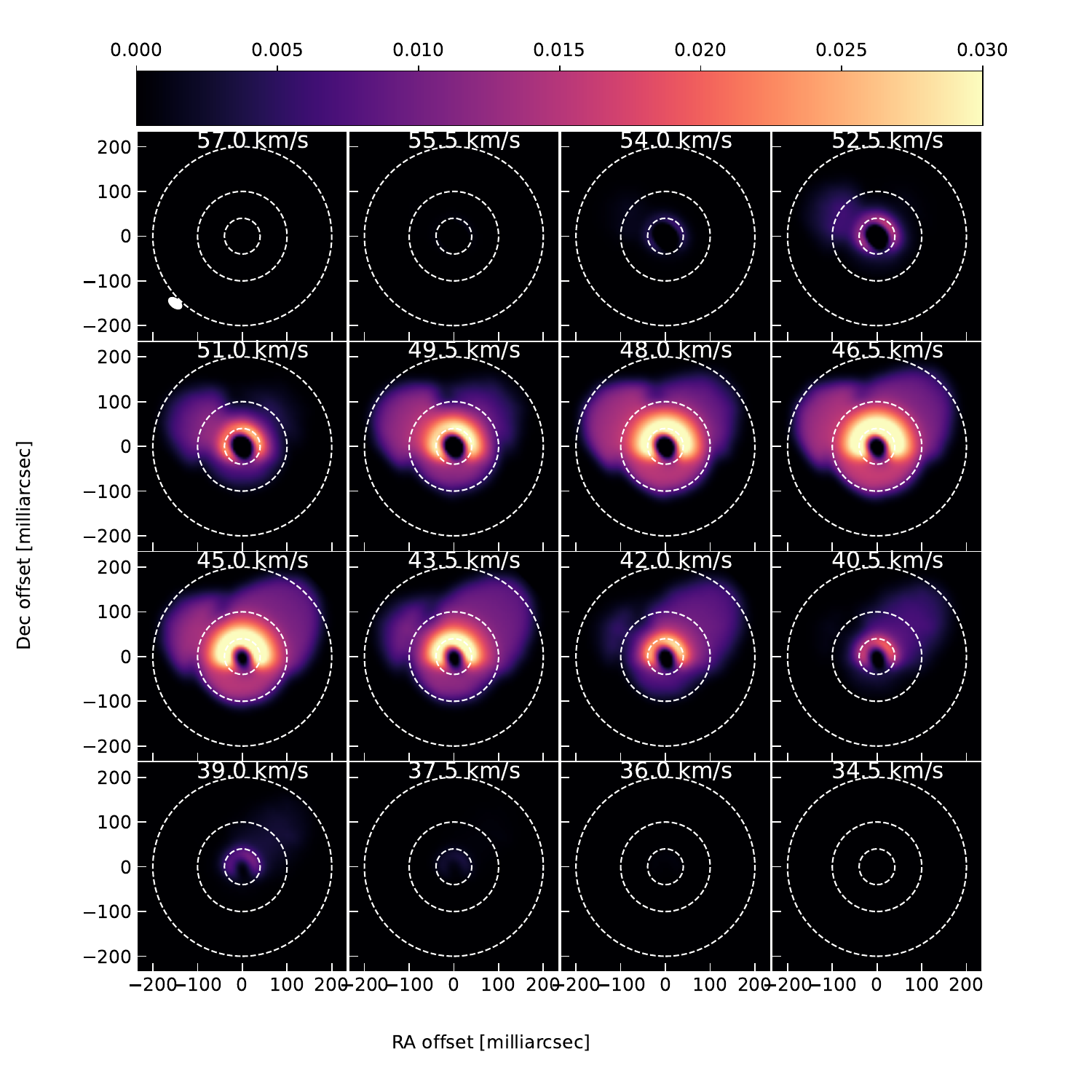}
        \caption{Channel maps of $^{13}$CO~$J=3-2$ emission from the radiative transfer model.
        The white dashed circles with radii 40, 100, and 200~mas indicate the region from where spectra were extracted to be compared to the radiative transfer models.}
        \label{fig:13CO-channels}%
    \end{figure*}

   \begin{figure*}[h!]
        \centering
        \includegraphics[width=0.4\textwidth]{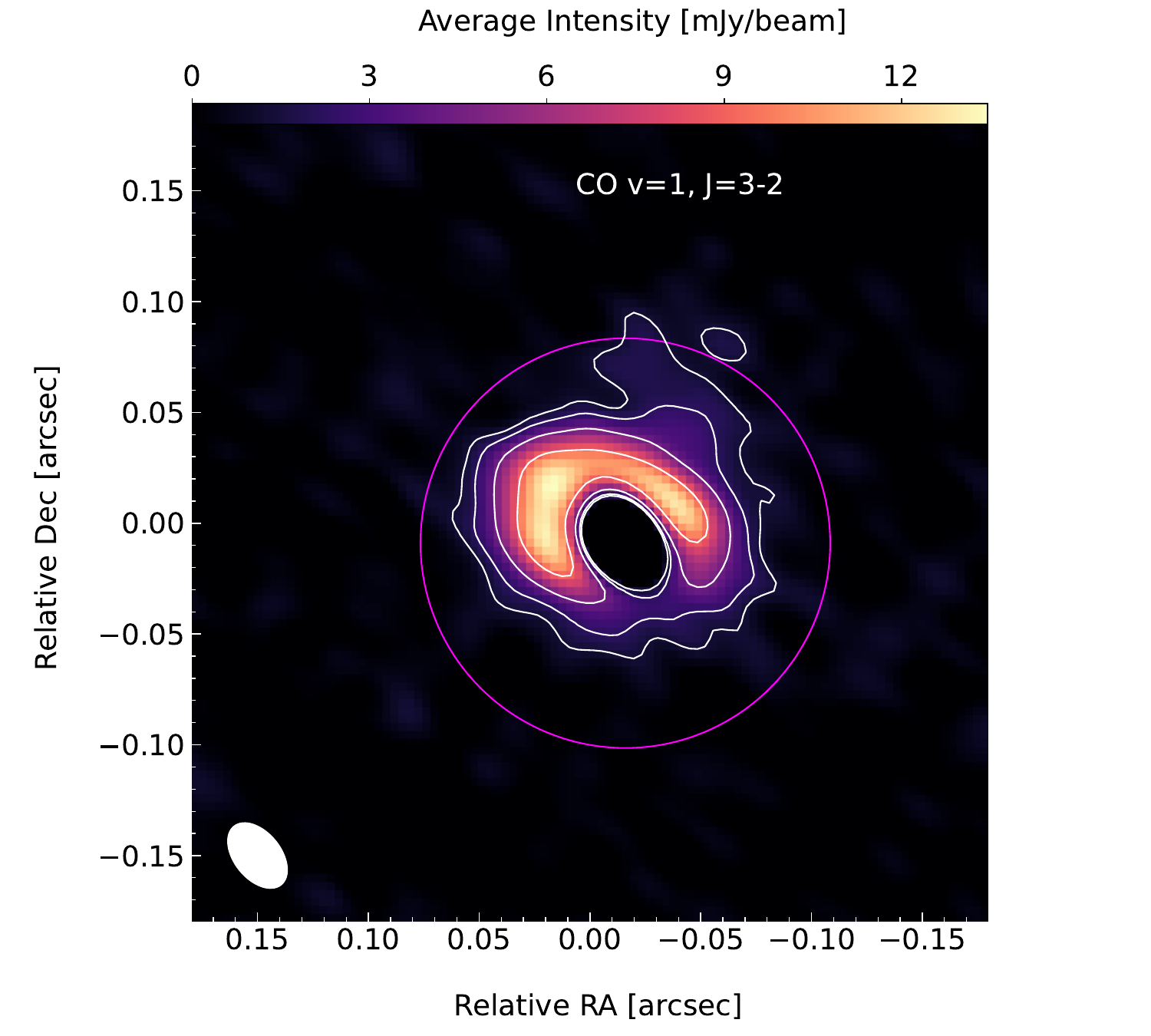}
        \caption{Observed map of CO~$v=1, J=3-2$ emission. The white contours mark emission at 3, 5, 10, and 20 times the value of the root mean square noise in the image ($0.45$~mJy/beam). The magenta circle indicates the size of the
        inner high-density region in the radiative transfer models.}
        \label{fig:COv1}%
    \end{figure*}

   \begin{figure*}[h!]
        \centering
        \includegraphics[width=0.9\textwidth]{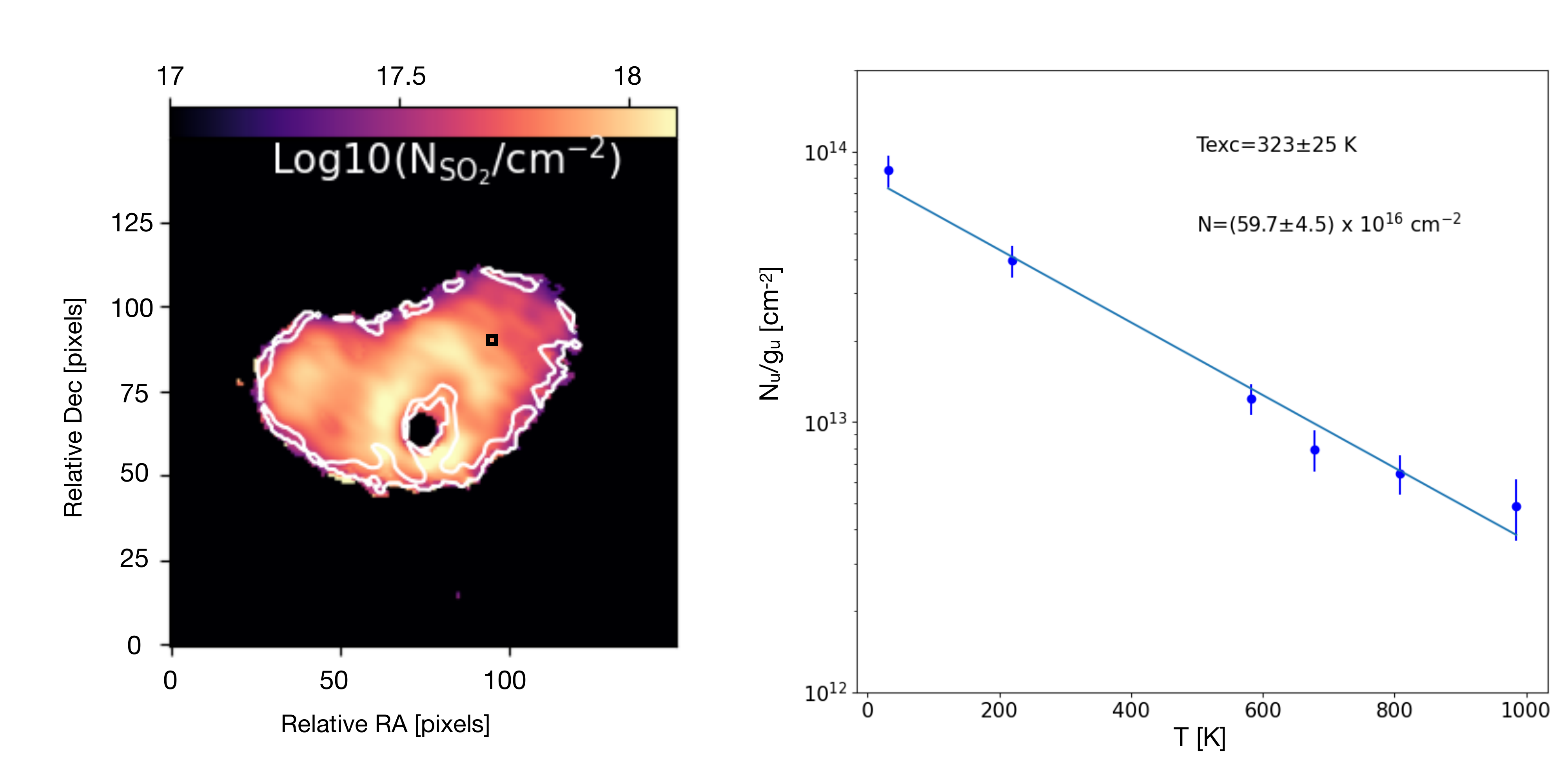}
        \caption{Example of population diagram obtained from a typical pixel. The left panel shows the SO$_2$ column density map (also shown in Fig.~\ref{fig:SO2_pop}) and the right panel shows the population diagram constructed
        using the line emission extracted from the selected pixel (indicated by the black square in the left panel).}
        \label{fig:popDiamExam}%
    \end{figure*}

\end{appendix}

\end{document}